\documentclass[prd,superscriptaddress,nofootinbib,11pt, tightenlines]{revtex4-2}

\usepackage{amsmath, amssymb, amsthm, graphicx, epsfig, fancyhdr,epsfig}
\usepackage[normalem]{ulem}
\usepackage{tikz-feynman}
\tikzfeynmanset{compat=1.1.0}
\usepackage{tikzsymbols}
\usepackage{tikz,xcolor,hyperref}
\usepackage{float}
\usepackage{xcolor}

\newcommand{\be}{\begin{equation}}
\newcommand{\ee}{\end{equation}}
\newcommand{\bea}{\begin{eqnarray}}
\newcommand{\eea}{\end{eqnarray}}

\def\d{{\rm d}}

\graphicspath{{Figures/}}

\begin{document}

\title{Cosmic inflation and $(g-2)_\mu$ in minimal gauged $L_\mu-L_\tau$ model}

\author{Arnab Paul}
\email{arnabpaul9292@gmail.com  }
\affiliation{School of Physical Sciences, Indian Association for the Cultivation of Science, Raja Subodh Chandra Mallick Rd, Jadavpur, Kolkata-700032, India}

\author{Sourov Roy}
\email{tpsr@iacs.res.in}
\affiliation{School of Physical Sciences, Indian Association for the Cultivation of Science, Raja Subodh Chandra Mallick Rd, Jadavpur, Kolkata-700032, India}

\author{Abhijit Kumar Saha}
\email{abhijit.saha@iopb.res.in}
\thanks{Corresponding author}
\affiliation{Institute of Physics, Sachivalaya Marg, Bhubaneswar-751005, India }

\begin{abstract}

{\bf Abstract:\,\,}The minimal $U(1)_{L_\mu-L_\tau}$ gauge symmetry extended Standard Model (SM) is a well motivated framework that resolves the discrepancy between the theoretical prediction and experimental observation of muon anomalous magnetic moment. We envisage the possibility of identifying the beyond Standard Model Higgs of $U(1)_{L_\mu-L_\tau}$ sector, non-minimally coupled to gravity, as the inflaton in the early universe, while being consistent with the $(g-2)_\mu$ data. Although the structure seems to be trivial, we observe that taking into consideration of a complete cosmological history starting from inflation through the reheating phase to late-time epoch along with existing constraints on $U(1)_{L_\mu-L_\tau}$ model parameters leave us a small window of allowed reheating temperature. This further results into restriction of $(n_s-r)$ plane which is far severe than the one in a generic non-minimal quartic inflationary set up.
\end{abstract}

\maketitle


\vspace{-1cm}
\section{Introduction}

The observed discrepancy between the measured and the theoretically predicted value of muon anomalous magnetic moment $(g-2)_\mu$ prompts us to look for beyond Standard Model (SM) physics. Such disagreement was previously reported by Brookhaven National Laboratory (BNL) E821 experiment \cite{Muong-2:2006rrc} and also recently measured by Fermilab E989 experiment \cite{Muong-2:2021ojo,fermi2023}. 
A precise analysis of the combined experimental data reveals \cite{Muong-2:2021ojo,fermi2023},
\begin{align}
    \Delta a_\mu=a_\mu^{\rm exp}-a_\mu^{\rm SM}=(249\pm 48)\times 10^{-11},
\end{align}
where we define $a_\mu=\left(\frac{g-2}{2}\right)_\mu$. The observed value of $a_\mu$ deviates from the SM predicted value \cite{Aoyama:2020ynm} at 
$5.0\sigma$ confidence level \footnote{Recently, Budapest-Marseille-Wuppertal (BMW) collaboration \cite{Borsanyi:2020mff} has claimed to find $\Delta a_\mu$ (lattice) $=107(69)\times 10^{-11}$ using QCD lattice simulations. This may bring down the discrepancy of the combined BNL-FNAL result with the SM within $2\sigma$. Also, if the hadronic vacuum polarisation contribution in the theoretical calculation of $a_\mu^{\rm SM}$ is taken from $e^{+}~e^{-}\rightarrow \pi^{+}~\pi^{-}$ cross section of CMD-3 experiment \cite{CMD-3:2023alj}, which itself is in tension with other $e^{+}~e^{-}$ data, the discrepancy of $\Delta a_\mu$ may decrease.}.

The minimal gauged $U(1)_{L_\mu-L_\tau}$ model, being anomaly free \cite{He:1990pn,He:1991qd} provides excellent scope to address the observed excess of the $(g-2)_\mu$ \cite{Baek:2001kca,Ma:2001md,Baek:2015fea,Heeck:2011wj,Banerjee:2020zvi,Sakurai:2022hwh,Zhou:2022cql}\,\footnote{See \cite{Lindner:2016bgg} for a broad review detailing various particle physics models accommodating $(g-2)_\mu$ data. A qualitatively alternate solution to the $(g-2)_\mu$ anomaly could be due to the presence of a new long-range force acting on spins \cite{Agrawal:2022wjm,Davoudiasl:2022gdg}, without the help of quantum corrections.}. It has been argued in \cite{Greljo:2022dwn} that, phenomenologically $X=L_\mu-L_\tau$ in a $U(1)_X$ gauged model is one of the very few preferred choices to explain the $(g-2)_\mu$ data. In the minimal version of $U(1)_{L_\mu-L_\tau}$ gauge theory, only second and third generations of SM leptons are charged under $U(1)_{L_\mu-L_\tau}$ symmetry and one also has a SM gauge singlet scalar. The gauge singlet scalar receives a non-zero vacuum expectation value (vev) which spontaneously breaks the $U(1)_{L_\mu-L_\tau}$ symmetry.  It is also possible to extend the minimal gauged $U(1)_{L_\mu-L_\tau}$ model by three right handed (RH) neutrinos without introducing any gauge anomaly, in order to accommodate correct order of neutrino mass and mixings, satisfying the neutrino oscillation data via standard seesaw mechanism \cite{Asai:2017ryy}. In this work, we adhere to the minimal gauged $U(1)_{L_\mu-L_\tau}$ model.

In another front, the cosmic microwave background (CMB) data reveals that our Universe is
spatially flat, homogeneous, and isotropic \footnote{Some recent works \cite{Secrest:2022uvx,Yeung:2022smn,Fosalba:2020gls,Aluri:2022hzs} have reported to find deviation from the cosmological principles related to homogeneity and isotropy of the Universe, however need further investigation.} which remain unexplained in the standard description of big bang cosmological theory. To interpret these shortcomings of the big bang cosmology, theory of cosmic inflation is postulated \cite{Guth:1980zm,Linde:1981mu,Starobinsky:1980te,Albrecht:1982wi}. During inflation, the Hubble parameter of the Universe remains almost constant, driving the Universe to undergo a phase of exponential expansion. Cosmic inflation also generates nearly scale invariant primordial scalar and tensor perturbations at super-horizon scale which seed formation of large scale structures while re-entering the horizon at a later stage.

Linking an inflationary scenario with a well motivated particle physics framework is an interesting exercise to pursue. There exists numerous works in this direction considering various types of supersymmetric and non-supersymmetric extensions of the SM of particle physics. The simplest possibility is to identify the SM Higgs as inflaton \cite{Bezrukov:2013fka}, the field responsible for the cosmic inflation. However, the tree level Higgs potential fails to provide a successful inflationary scenario and one requires to introduce non-minimal coupling ($\xi$) of the SM Higgs with gravity to make the SM Higgs inflation compatible with experimental data. It turns out that one needs a large $\xi\sim \mathcal{O}(10^4)$ to be in agreement with the observed curvature perturbations \cite{Bezrukov:2007ep}, which gives rise to the question of naturalness from unitarity point of view \cite{Barbon:2009ya,Lerner:2009na}. This issue can be possibly eliminated if the inflaton is identified with a gauge singlet scalar. Presence of gauge singlet scalar is common in different Abelian gauge extended beyond standard model (BSM) frameworks.

\begin{figure}[t]
    \centering
    \includegraphics[height=6cm,width=11cm]{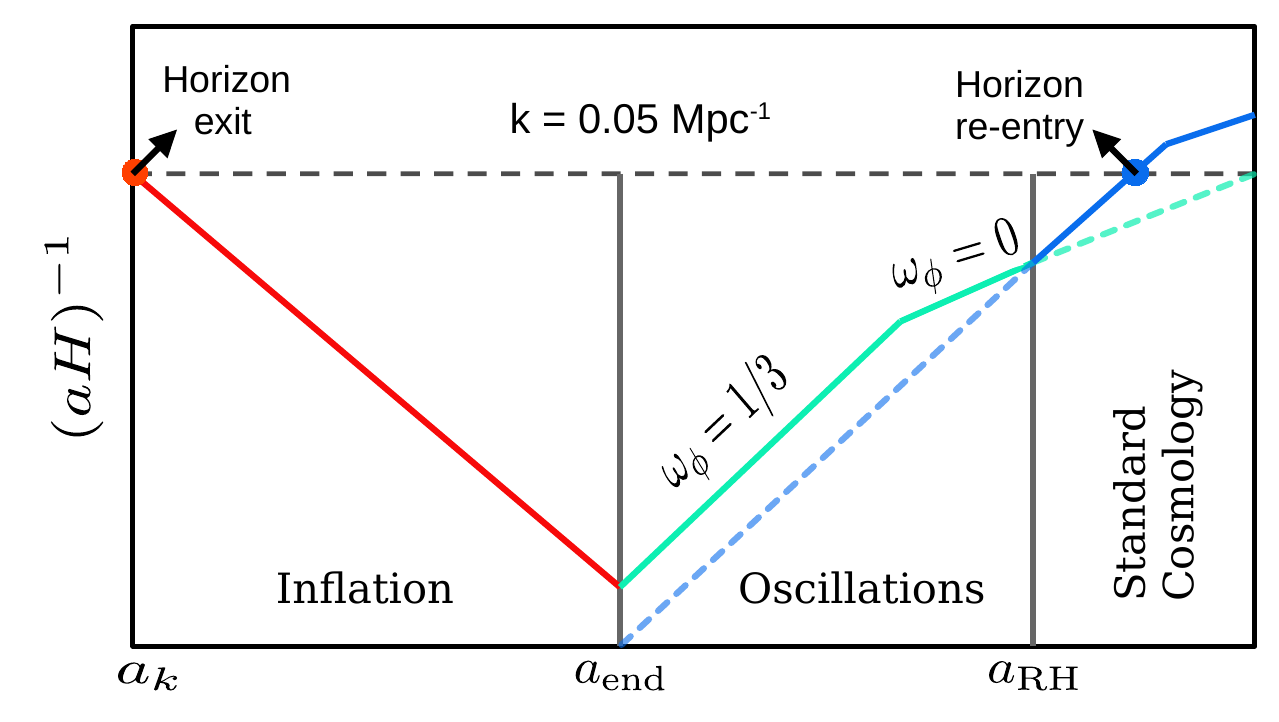}
    \caption{The evolution of comoving Hubble radius $\frac{1}{(aH)}$ is shown as function of a scale factor of the Universe. The red colored line indicates the evolution of comoving Hubble radius during inflation that exited the horizon at $a=a_k$, for a particular number of inflationary e-fold $N_{\rm e}$. The inflation ends at $a=a_{\rm end}$ and the evolution of $\frac{1}{(aH)}$ during inflaton oscillation era is indicated by solid cyan colored line. At $a=a_{\rm RH}$, the reheating of the Universe is completed and afterwards the RD universe prior to BBN sets in. The solid blue line marks the evolution of $\frac{1}{(aH)}$ for a standard observable Universe, extrapolated up to $a=a_{\rm RH}$.}
    \label{fig:evolution}
\end{figure}

In this work, we analyse the dynamics of cosmic inflation in the minimal $U(1)_{L_\mu-L_\tau}$ framework where the SM gauge singlet scalar is identified with the inflaton. We have adopted the conventional inflationary set up with non-minimal coupling of inflaton to gravity \cite{Park:2008hz,Okada:2010jf,Bezrukov:2013fca,Linde:2014nna} in order to satisfy the cosmological inflationary constraints as provided by Planck+BICEP/Keck \cite{Planck:2018jri,BicepKeck:2021ybl}\footnote{Note that, the cosmic inflation in minimal gauged $U(1)_{L_\mu-L_\tau}$ model was also discussed in Ref.\cite{Asai:2020qax}, however the inflation sector was different from the one presented in this paper.}.  In particular we examine the viability of $(g-2)_\mu$ satisfying parameter space in providing a consistent cosmological scenario from the onset of primordial inflation to the end of matter-radiation equality with the intermediate reheating phase\,\cite{Albrecht:1982mp,Dolgov:1989us,Kofman:1994rk,Shtanov:1994ce}. We observe that the measurements of $(g-2)_\mu$ along with other observational constraints restrict the $U(1)_{L_\mu-L_\tau}$ breaking scale to remain within $\sim (9-75)$\,GeV. This estimate of $U(1)_{L_\mu-L_\tau}$ breaking scale along with the inflationary constraints determine the inflaton mass. The inflaton mass is an important quantity which decides the dynamics of the inflaton oscillations and sets the kinematics of the energy transfer processes of inflaton to radiation. For a given particular inflation model with fixed number of inflationary e-folds the evolution of the Hubble horizon can be easily determined. On the other hand from observations, we know the dynamics of standard Hubble horizon as a function of scale factor from radiation dominated epoch till the present time. Therefore, to enter into a standard radiation dominated phase from the inflaton dominated Universe, the Hubble horizon during inflaton decay (reheating) must match the standard one before the onset of Big Bang Nucleosynthesis (BBN) \cite{Liddle:2003as,Martin:2010kz,Drewes:2017fmn}.
We present a schematic diagram of this evolution as a function of scale factor in Fig.\,\ref{fig:evolution}. This observation tells us the completion time (or scale factor) of inflaton decay which in turn fixes the inflaton interaction strength with the SM particles for a given number of inflationary e-fold.

In our case, the energy transfer of inflation sector to radiation bath occurs via perturbative decay of inflaton. {Depending on the mass scale, the inflaton can decay to various final state particles. For $m_\phi<m_{Z^\prime}$, where $m_\phi$ and $m_{Z^\prime}$ indicate the inflaton and $L_\mu-L_\tau$ gauge boson masses respectively, $\phi\to \overline{f} f$ decay ($f$ being a representative SM particles which are kinematically allowed) is the only possibility while other possible decay modes are kinematically forbidden. {In the other case, due to the presence of new decay modes having $Z^\prime$ in the final states, the inflaton decay happens at a rapid rate. In fact we found that the completion of inflation decay gets over even before the onset of $\omega_\phi=0$ phase (see Fig.\,\ref{fig:evolution}) due to the largeness of $U(1)_{L_\mu-L_\tau}$ gauge coupling. This leads to a direct transition of inflaton dominated Universe (with $\omega_\phi=1/3$) to standard radiation Universe ($\omega=1/3$) given the absence of $\omega_\phi=0$ phase.}
Subsequently, we study the correspondence of the inflationary parameters namely number of e-folds during inflation ($N_{\rm e}$) and non-minimal coupling of inflaton ($\xi$) with inflaton mass ($m_\phi$) and Higgs-singlet scalar mixing angle, $\theta$ that control the fate of the standard post-inflationary Universe for {$m_\phi<m_{Z^\prime}$ case}. Note that the $m_\phi$ - $\sin\theta$ plane is already constrained by various experimental observations in the sub-GeV inflaton mass range. The above mentioned correspondence between $N_e$ and $\theta$ corresponding to different $\xi$ values further add} new constraints on the inflaton mass\,-\,$\theta$ plane, mainly arising from the stability of the inflaton potential during inflation, positivity of e-foldings in different epochs and the completion of inflaton decay before BBN. We find that these new constraints along with existing ones on the inflaton mass\,-\,$\theta$ plane significantly restrict the amount of both $N_{\rm e}$ and $\xi$ and thereby leaving strong impact on the prediction of inflationary observables such as $n_s$ and $r$. {On the other hand for $m_\phi>m_{Z^\prime}$ case, we end up with an unique set of value for $(n_s,r)$ owing to the fact that $N_e$ turns independent of the reheating temperature. After combining the outcomes for these two viable cases in the present study, considering the minimal version of $L_\mu-L_\tau$ model, we have obtained the predictions for $n_s$ and $r$ in the parameter region which is in agreement with $(g-2)_\mu$ data.} Our results for ($n_s,r$) turn out to be very predictive and much more restrictive than the one in a generic non-minimal quartic inflationary set up.

The plan of the paper is as follows. In section \ref{secmodel} we discuss the $U(1)_{L_\mu-L_\tau}$ model briefly. In section \ref{method} we explain the methodology adopted in this work. Section \ref{reheat} describes the dynamics of the inflaton after inflation till reheating. In section \ref{results} and \ref{conclusion} we discuss the results and conclusions respectively.

\section{$L_\mu-L_\tau$ Model}\label{secmodel}
In the minimal $U(1)_{L_\mu-L_\tau}$ gauged model, the gauge anomalies get cancelled between $\mu$ and $\tau$ leptonic generations, even without the introduction of any additional chiral fermions. One also incorporates a SM gauge singlet scalar field ($\varphi$) that receives non-zero VEV and breaks the $U(1)_{L_\mu-L_\tau}$ gauge symmetry spontaneously. The scalar sector of the $U(1)_{L_\mu-L_\tau}$ gauged framework have following structure of the potential.
\begin{align}
V(|H|,|\varphi|) =  \lambda_\phi \left(|\varphi|^2 -\frac{v'^2}{2}\right)^2 + \lambda_H \left(|H|^2 -\frac{v^2_H}{2}\right)^2
    +  \lambda_{H\phi} \left(|H|^2 -\frac{v^2_H}{2}\right)\left(|\varphi|^2 -\frac{v'^2}{2}\right),\label{eq:scalarPot}
\end{align}
where we have assumed all dimensionful and dimensionless couplings as real and positive. We have assigned `+2' $U(1)_{L_\mu-L_\tau}$ charge to the $\varphi$ field. After the electroweak symmetry breaking (EWSB), considering unitary gauge, we write,
\begin{align}
H=\begin{pmatrix}
0\\
\frac{v_H+h}{\sqrt{2}} 
\end{pmatrix}
\textrm{~~~~and~~~~}
\varphi=\frac{1}{\sqrt{2}}(v'+\phi).
\end{align}
The scalar mass squared matrix contains off-diagonal terms and after EWSB, mixing between $h$ and $\phi$ takes place. The mass eigenstates ($\phi_1$ and $\phi_2$) are related to $h$ and $\phi$ as,
\begin{eqnarray}
\begin{bmatrix} h \\ \phi \end{bmatrix}   =
\begin{bmatrix} \cos\theta &   -\sin\theta \\ \sin\theta & \cos\theta  \end{bmatrix} \begin{bmatrix} \phi_1 \\ \phi_2 
\end{bmatrix},   \nonumber
\end{eqnarray}
where $\theta$ is the mixing angle between $h$ and $\phi$. The mixing angle $\theta$ and the masses $m_{\phi_1}$ and $m_{\phi_2}$ of the mass eigenstates $\phi_1$ and $\phi_2$ are then given by,
\begin{eqnarray}
\tan2\theta&=&\frac{2 v' v_{H}  \lambda_{H\phi}}{ m_h^2 -m_\phi^2}\\
m_{\phi_1}^2 &=& m_h^2 - \left(m_\phi^2 - m_h^2 \right) \frac{\sin^2\theta}{1-2 \sin^2\theta}\\
    m_{\phi_2}^2 &=& m_\phi^2 + \left(m_\phi^2 - m_h^2 \right) \frac{\sin^2\theta}{1-2 \sin^2\theta},
\end{eqnarray}
where $m_{\phi} = \sqrt{2 \lambda_\phi} v'$, $m_h = \sqrt{2 \lambda_{H}} v_{H}$. For small $\theta$, which is the scenario we are interested in, $\phi_1$ and $\phi_2$ are mostly the SM Higgs having mass 125 GeV and inflaton $\phi$ respectively. The mass of the mostly inflaton state $\phi_2$ is then $m_{\phi_2}\sim m_\phi$, whereas $m_{\phi_1}\sim m_h$.

In the broken phase of $SU(2)_L\times U(1)_Y\times U(1)_{L_\mu-L_\tau}$, we also write the following Lagrangian,
\begin{align}
    \mathcal{L}=\mathcal{L}_{\rm SM}-\frac{1}{4}Z^{\prime \alpha\beta}Z^\prime_{\,\alpha\beta}+\frac{m_{Z^{\prime}}^2}{2}Z^{\prime\alpha} Z^\prime_{\alpha}+Z^\prime_{\alpha}J_{\mu\tau}^\alpha,
\end{align}
where $Z^{\prime}_{\alpha\beta}=\partial_\alpha Z^\prime_\beta-\partial_\beta Z^\prime_{\alpha}$ is the field strength tensor for the $U(1)_{L_\mu-L\tau}$ gauge symmetry, $m_{Z^\prime}=2g^\prime v^\prime$ represents the mass of the new gauge boson $Z^\prime$. The $J_{\mu\tau}^\alpha$ stands for the $\mu-\tau$ current as expressed as,
\begin{align}
    J_{\mu\tau}^\alpha=g^\prime(\overline{\mu}\gamma^\alpha\mu+\overline{\nu_\mu}\gamma^\alpha P_L\nu_\mu-\overline{\tau}\gamma^\alpha\tau-\overline{\nu_\tau}\gamma^\alpha P_L\tau_\mu),
\end{align}
where $P_L=\frac{1}{2}(1-\gamma_5)$ is the left chiral operator and $g^\prime$ stands for the $U(1)_{L_\mu-L_\tau}$ gauge coupling. Although the tree level kinetic mixing is assumed to be zero, at loop level,  $Z^\prime$ mixes with the photon that leads to coupling of $Z^\prime$ with electrons \cite{Kamada:2015era,Araki:2017wyg}.

The ability of minimal $U(1)_{L_\mu-L_\tau}$ framework to accommodate the experimentally observed value of $(g-2)_\mu$ has long been recognized. 
This contribution to the anomalous magnetic moment of the muon ($\Delta a_\mu$) arises dominantly from the loop involving the $Z^\prime$ boson, which can be expressed as follows \cite{Baek:2001kca,Ma:2001md}:
\begin{align}
    \Delta a_\mu= \frac{{g^\prime}^2}{4\pi^2}\int_0^1dx\,\frac{x(1-x)^2}{(1-x)^2+r x},
\end{align}
where $r=\left(\frac{m_{Z^\prime}}{m_\mu}\right)^2$, with $m_\mu$ being the $\mu$ mass. 
\begin{figure}
    \centering
    \includegraphics[height=8cm,width=10cm]{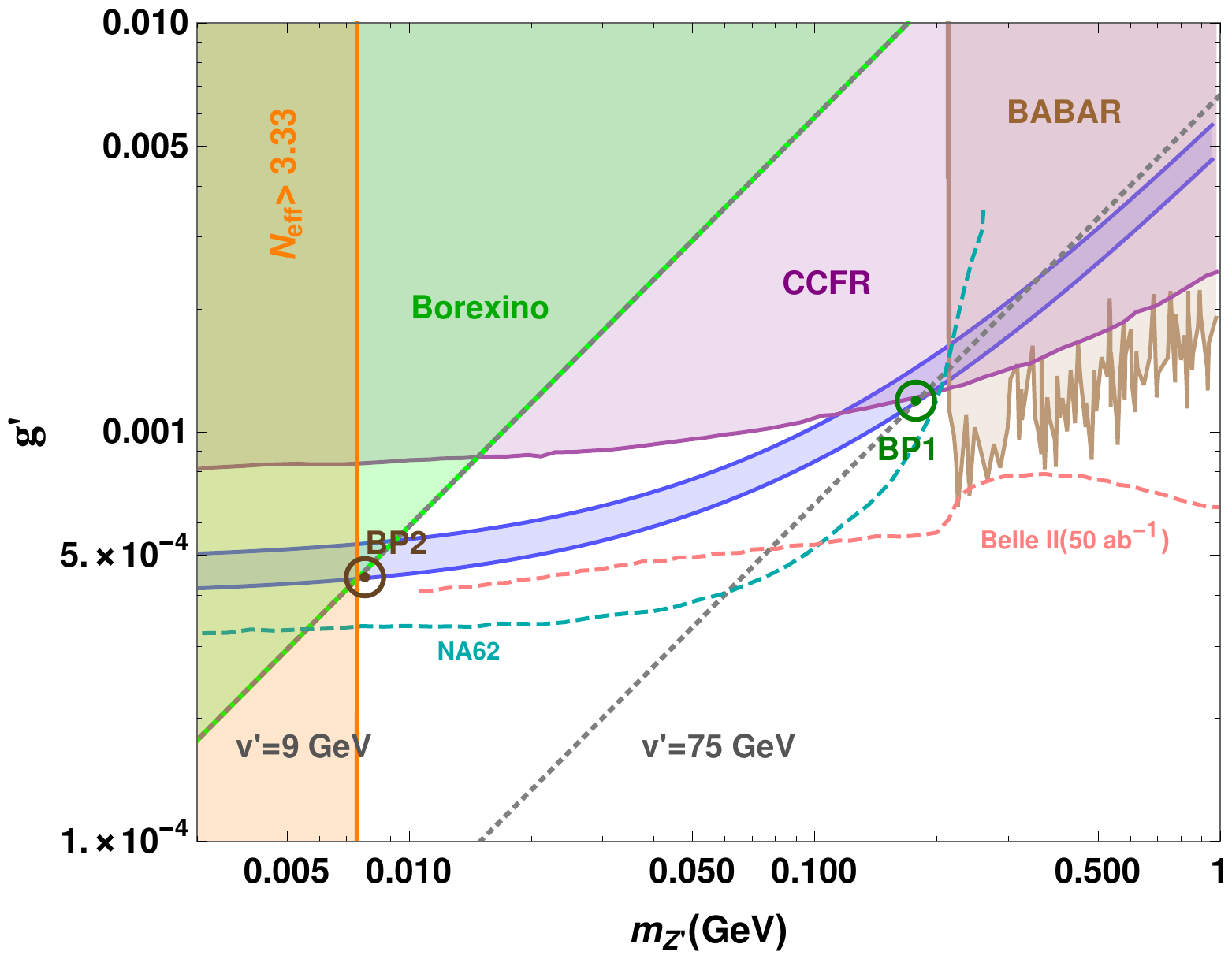}
    \caption{Existing constraints on the $m_{Z'}-g'$ plane is shown. The region in blue color successfully accommodates the $(g-2)_\mu$ data. Constraints from BABAR\,\cite{BaBar:2016sci}, Borexino\,\cite{Greljo:2022dwn,Borexino:2013zhu} and CCFR\,\cite{PhysRevLett.66.3117} are highlighted with green, brown and purple colors respectively. The two dashed lines correspond to $v^\prime=9$\,GeV and $v^\prime=75$\,GeV respectively. {The orange shaded region ($m_\phi\lesssim 7.4$\,MeV \cite{Kamada:2018zxi}) is disallowed by the Planck data\,\cite{Planck:2018jri} that rules out $N_{\rm eff}>3.33$ at 2$\sigma$}. The upcoming sensitivities of NA62 and Belle II are taken from \cite{Krnjaic:2019rsv} and \cite{Belle-II:2022cgf} respectively.}
    \label{fig:mZP-gP plane}
\end{figure}
In Fig.\,\ref{fig:mZP-gP plane}, the region, highlighted in blue represents the parameter space where the experimentally preferred values can be accommodated successfully. The stringent constraints on $m_{Z'}-g'$ plane appear due to neutrino interaction of $Z^\prime$ at CCFR \cite{PhysRevLett.66.3117} and solar neutrino scattering at Borexino \cite{Borexino:2013zhu}. Additionally four $\mu$-final state searches at BABAR \cite{BaBar:2016sci} {and bound on effective number of neutrino degrees of freedom, $N_{\rm eff}$ \cite{Kamada:2018zxi} from Planck experiment \cite{Planck:2018jri}} strongly restrict the $m_{Z'}-g'$ plane. Future observations from NA62 \cite{Krnjaic:2019rsv} and Belle\,II\,(50 ab$^{-1}$) \cite{Belle-II:2022cgf} can potentially rule out the whole $(g-2)_\mu$ satisfying parameter space. Taking into account all the relevant existing constraints we infer that present experimentally preferred value of $\Delta a_\mu$ requires $9\,{\rm GeV}\lesssim v^\prime\lesssim 75\,{\rm GeV}$ as depicted by two black dotted line in Fig.\,\ref{fig:mZP-gP plane}.

{To note, in Fig.\,\ref{fig:mZP-gP plane} (and also in Figs.\,\ref{fig:mphi_sinthBP1} and \ref{fig:mphi_sinthBP2}) we have not considered the possible impact of neutrino self-interactions mediated by MeV scale $Z^\prime$ in the physics of supernova neutrinos. It is discussed in \cite{Kamada:2015era,Kamada:2018zxi} that non-standard neutrino self-interactions mediated by $Z^\prime$ in the present framework may delay the diffusion time of $\nu_\mu$ and $\nu_\tau$ which is strictly constrained by observation of neutrino burst in SN1987a\,\cite{PhysRevLett.58.1490,PhysRevLett.58.1494}. This supernova event can potentially pose challenge to the available $(g-2)_\mu$ satisfying parameter space in the $m_{Z^\prime}-g^\prime$ plane of Fig.\,\ref{fig:mZP-gP plane}. It was also suggested in \cite{Kamada:2015era,Kamada:2018zxi} that such a delay in diffusion can be compensated if there exists an additional cooling mechanism present, which in our scenario could be possible due to the presence of SM gauge singlet $\phi$ depending on its production rate at supernova core and subsequently energy loss dynamics. However, there are large uncertainties involved in the calculation of diffusion time of neutrinos in presence of non-standard interactions \textit{e.g.} supernova modelling, limited statistics of observed events {\it etc.} as reported in \cite{Kamada:2015era}. 
In this context, we would like to mention that an earlier study in Ref.\,\cite{MANOHAR1987217} also suggested that the neutrino non-standard self-interactions may influence supernova cooling, however, such observation is challenged by Ref.\,\cite{DICUS198984} and more recently by Ref.\,\cite{Cerdeno:2023kqo}, suggesting that non-standard neutrino self-interactions have very limited impact on supernova burst.
Thus, an improved and precise analysis on the possible impact of $Z^\prime$ mediated non-standard neutrino self-interaction on SN1987a is required before arriving at a conclusion, which can possibly be pursued in a separate study.}

\section{The proposed methodology}\label{method}
At the onset of inflation, we consider the field $\phi$ as the sole species to remain abundant in the early Universe. The action in that case is given by,
\begin{align}
    S = \int \d^4 x \sqrt{-g} \left[-\frac{M_P^2}{2}\left(1+\frac{\xi \phi^2}{M_P^2}\right) R + {1\over 2} g^{\mu \nu}\partial_\mu\phi \, \partial_\nu\phi-V(\phi) \right].
\end{align}
Here $V(\phi)$ is the inflationary potential, expressed as 
\begin{equation}
    V\supset \frac{\lambda_\phi}{4}\phi^4,
\end{equation}
with the role of the inflaton being served by the CP even neutral scalar $\phi$. Here $R$ is the Ricci scalar and $\xi$ stands for the non-minimal coupling of $\phi$ to gravity. Note that the non-minimal interaction of $\phi$ with gravity originally arises from $\xi|\varphi|^2 R$ term and hence does not violate the $U(1)_{L_\mu-L_\tau}$ symmetry. Since the gravity is non-minimal here, a conformal transformation of metric is essential in order to preserve the validity of Einstein Hilbert gravity. The metric transformation is given by,
\begin{align}
\bar g_{\mu\nu}=\Omega^2 g_{\mu\nu},~\text{~with~}~\Omega^2=1+\frac{\xi \phi^2}{M_P^2}
\end{align}
To get rid of the non-canonical kinetic term, induced by the metric transformation, following transformation of $\phi$ is also required,
\begin{align}\label{eq:fieldTrans}
    \frac{\d \chi}{\d \phi}=\frac{1}{\Omega^2}\sqrt{\Omega^2+6\xi^2 \frac{\phi^2}{M_P^2}}
\end{align}

{With these, we finally obtain the inflationary potential in the Einstein frame as defined by,
\begin{align}
    V_E(\chi)=\frac{1}{\Omega^4(\phi(\chi))}V_J(\phi(\chi))
\end{align}
where $V_J(\phi)=\frac{\lambda_\phi}{4}\phi^4$, the potential in the Jordan frame.
Next we define the slow roll parameters $\epsilon$ and $\eta$ as,
\begin{eqnarray}
\epsilon=\frac{ M_{P}^2}{2} \left(\frac{V_E'}{V_E}\right)^2, \; \; 
\eta=
M_{P}^2\left(\frac{V_E''}{V_E}\right)\,,
\label{eq:SRCond}
\end{eqnarray}
where $V_E^\prime=\frac{\partial V_E}{\partial \chi}$ and $M_P$ is the reduced Planck scale.
The number of e-folds during inflation $N_{\rm e}=
{\rm log}\left(\frac{a_{\rm end}}{a_*}\right)$ is given by,
  \begin{eqnarray}\label{eq:Ne}
N_{\rm e}=\frac{1}{M_{P}^2}\int_{\chi_{\rm end}}^{\chi_*}\frac{V_E }{V_E^\prime} d\chi,
 \label{eq:EFold}
\end{eqnarray} 
where $\chi_*$ and $\chi_{\rm end}$ indicates the inflaton field values at horizon exit and end of inflation. $\chi_{\rm end}$ can be simply found by using the condition max$[\epsilon,\eta]=1$ that dictates the violation of slow roll of inflaton field. The above relation connects the number of e-fold during inflation with the inflaton field value at horzion exit. The inflationary scalar power spectrum $P_s$ is expressed as, 
\begin{equation}
P_s(k)=A_s \left(\frac{k}{k_0}\right)^{n_s-1},
\end{equation}
where $k_0 = 0.05$ Mpc$^{-1}$ is the pivot scale. The amplitude $A_s=2.2\times 10^{-9}$\,\cite{Planck:2018jri} is experimentally well measured and $n_s$ is the spectral index. In terms of the scalar potential during inflation they take the following form,
\begin{equation} 
A_s= \frac{1}{24 \pi^2 M_P^4}\left. \frac{V_E}{ \epsilon } \right|_{k_0}, \; \; n_s = 1-6\epsilon+2\eta. 
 \label{PSpec}
\end{equation}
Under the assumption of slow roll, the tensor to scalar ratio $r$ is defined as, 
\begin{equation}\label{eq:rdef}
    r = 16 \epsilon .
\end{equation}
Note that, the predictions for the quantities $A_s,n_s$ and $r$ have to be computed at horizon exit value of the inflaton field. After inflation, $\chi\ll\frac{M_P}{\sqrt{\xi}}$ and in this limit $\chi$ can be safely identified as $\phi$ following Eq.(\ref{eq:fieldTrans}).}

It should be noted that the current setup under consideration leads to a high-scale inflationary scenario with $\chi> M_P$ (implying $\phi>M_P$ as well) during inflation. This means that the inflation has occurred at energy scales significantly higher than the EWSB scale and hence it is safe to ignore the $\frac{1}{2}\lambda_\phi v'^2|\phi|^2$ term from Eq.(\ref{eq:scalarPot}) during inflation. The inflaton mass at its minimum is given by $m_\phi\simeq m_{\phi_2}$ in the small $\sin\theta$ limit. A generic quartic inflationary potential \cite{Linde:1983gd} {\it i.e.} when $\xi=0$, is disfavored by the Planck-2018 constraints \cite{Planck:2018jri} due to the prediction of large tensor to scalar ratio $r\sim 0.133$ \cite{Linde:1983gd}. The nonminimal coupling of inflaton to gravity assists into flattening of inflationary potential that further leads to the revival of a quartic inflationary set up. In this modified set up of quartic inflation, one can simply estimate the values of spectral index ($n_s$), $r$ and scalar curvature perturbation spectrum $A_s$, under the slow roll approximation and using the conventional approach \cite{Okada:2010jf}. The observed value of $A_s$ is $2.2\times 10^{-9}$ \cite{Planck:2018jri}, which uniquely constrains the parameter $\lambda_\phi$ as a function of $\xi$. The predictions of $n_s$ and $r$ have no dependence on $\lambda_\phi$.

It is worth mentioning that inflection point inflation scenarios in a few Abelian gauge ({\it e.g.} $U(1)_{B-L}$ \cite{Okada:2016ssd}) extended models have been widely studied, which does not require the non-minimal coupling to gravity in order to flatten the quartic inflationary potential. In these scenarios, interplay of the quantum corrections from the BSM gauge coupling and the Yukawa couplings give rise to the required flatness. However, the current model being a minimal one, does not have any additional Yukawa couplings. 

\begin{figure}[h]
    \centering
    \includegraphics[height=6.5cm,width=14cm]{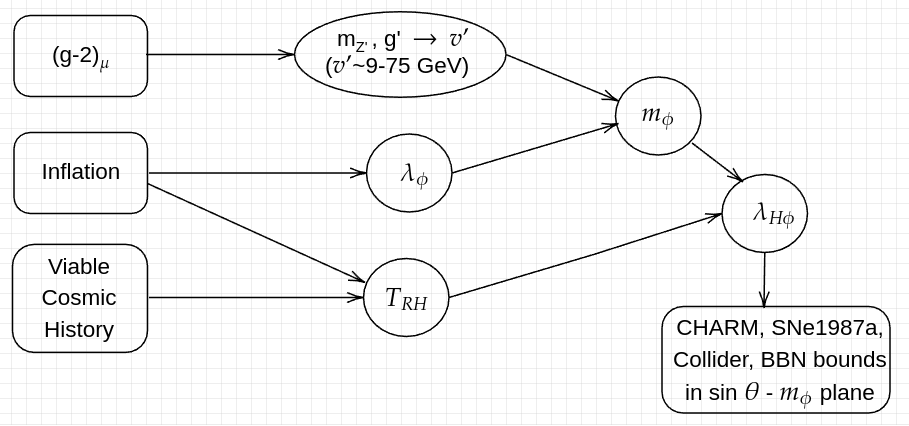}
    \caption{\label{flow} A schematic representation of the strategy to connect $(g-2)_\mu$ data with the cosmological predictions for $(n_s,r)$. {This pictorial representation (particularly the reheating part) is valid for $m_\phi<m_Z^\prime$ regime. For the other case, the reheating dynamics will depend on additional gauge
parameters as well.}}
    \label{fig:cartoon2}
\end{figure}

Next we describe our adopted methodology (as summarised in Fig.\,\ref{fig:cartoon2}) that reveals how the satisfaction of $(g-2)_\mu$ data impact the predictions of the inflationary observables $(n_s,r)$. As earlier mentioned, a consistent $\Delta a_\mu$ data requires $9\,{\rm GeV}\lesssim v^\prime\lesssim 75\,{\rm GeV}$. On the other hand, the successful dynamics of inflation in generating the observed value of scalar curvature spectrum \cite{Planck:2018jri} uniquely fixes $\lambda_\phi$ as a function of the non-minimal coupling parameter $\xi$. Choosing a particular $v^\prime$ within its preferred range along with $\lambda_\phi(\xi)$ provide us the dependence of the inflaton mass ($m_\phi)$ on $\xi$. Now we consider the inflaton to decay into radiation perturbatively via Higgs portal 
\footnote{The impact of preheating at the early stage of inflaton oscillation era leave minimal impact in our final results, as will be justified shortly.}. 
{Recall that previously we commented that the $m_\phi>m_Z^\prime$ regime is not compatible with dynamics of post inflationary observed Universe and thus we focus on $m_\phi<m_Z^\prime$ case to describe the methodology here. In section\,\ref{sec:resCII} we demonstrate the inability of $m_\phi>m_Z^\prime$ region to accommodate a consistent and unified picture of inflation\,-\,reheating\,-\,present Universe. For} $m_\phi<m_Z^\prime$ case, the reheating temperature of the Universe ($T_{\rm RH}$) turns out to be a function of $m_\phi$ and $\lambda_{H\phi}$. Now for a given number of inflationary e-fold ($N_{\rm e}$), the $T_{\rm RH}$ of the Universe cannot be arbitrary. This means for a fixed $\xi$ (or $m_\phi$), the $\lambda_{H\phi}$ has be to uniquely fixed as well. It is known that the $m_\phi-\lambda_{H\phi}$ plane is severely constrained by different existing experiments namely CHARM \cite{BERGSMA1985458}, SN1987a \cite{Dev:2020eam,Krnjaic:2015mbs}, BBN {\it etc}. On the top of that, there are additional constraints arising to ensure the stability of inflaton potential during inflation and positivity of number of e-fold during inflaton oscillation and reheating. Such constraints in the $m_\phi-\lambda_{H\phi}$ plane strongly restricts $N_{\rm e}$ as a function of $\xi$ which in turn poses constraint in the $n_s-r$ plane.


\section{Post-Inflationary dynamics and Reheating}\label{reheat}

After the slow-roll conditions get violated, the inflation ends and the inflaton begins to oscillate around its minimum. Initially the amplitude of the oscillation is large and the potential remains almost quartic and at later stage the quadratic term of $\phi$ starts\,\footnote{In the Einstein frame, $\chi$ is the canonical field during inflation, connected with $\phi$ via Eq.(\ref{eq:fieldTrans})} to dominate. It is well known that the average equation of state (e.o.s) parameter during the oscillation regime is $\omega=\frac{n-2}{n+2}$ for a $\phi^n$ potential. Hence at initial phases of inflaton oscillation, $\omega=\frac{1}{3}$ and subsequently $\omega=0$ in the $\phi^2$ dominated phase. The transition of {\it e.o.s} parameter from $\frac{1}{3}$ to $0$ can be noticed in Fig.\,\ref{fig:evolution} as well. Importantly, the maximum oscillation amplitude of the inflaton at the point of transition from radiation-like phase ($\omega=\frac{1}{3}$) to matter-like phase $w=0$ is function of the ratio $\frac{m_\phi}{\sqrt{\lambda_\phi}}$ which is nothing but $\sqrt{2}v^\prime$.

Now, at Hubble horizon exit for a comoving mode $k$, we can write $k=a_k H_k$ with $a_k$ (same as $a_*$ in the definition of $N_e$) and $H_k=\frac{\pi M_P \sqrt{r A_s}}{\sqrt{2}}$ being the scale factor and Hubble rate during the horizon exit of mode $k$. Here we consider $k=k_{\rm pivot}^{\rm Planck}=0.05~\rm Mpc^{-1}$. Then it follows,
\begin{equation}
    \ln\left(\frac{k}{a_k H_k}\right)=\ln\left(\frac{a_{\phi,\rm rad}}{a_k}\frac{a_{\phi,\rm matter}}{a_{\phi,\rm rad}}\frac{a_{\rm RH}}{a_{\phi,\rm matter}}\frac{a_{k,\rm late}}{a_{\rm RH}}\frac{k}{a_{k,\rm late} H_k}\right)=0,
    \label{eq:k=aH}
\end{equation}
where $a_{\phi,\rm rad}$ (same as $a_{\rm end}$) denote the scale factor of the Universe at the beginning of inflaton oscillation (at the end of inflation) with e.o.s. $\frac{1}{3}$ and $a_{\phi,\rm matter}$ stands for the scale factor at the cross-over where inflaton becomes matter like having $\omega=0$. $a_{\rm RH}$ corresponds to the scale factor at the end of reheating and $a_{k,\rm late}\sim 4.5\times10^{-5}$ indicates the scale factor when $k=0.05~\rm Mpc^{-1}$ mode re-enters the Hubble horizon. The Eq.(\ref{eq:k=aH}) can be further translated to,
\begin{align}
    N_{\rm e}+{\rm ln}\left(\frac{\rho_{\phi,\rm rad}}{\rho_{\phi,\rm matter}}\right)^{\frac{1}{4}}+{\rm ln}\left(\frac{\rho_{\phi,\rm matter}}{\rho_{\phi,\rm RH}}\right)^{\frac{1}{3}}+{\rm ln}\left(\frac{\rho_{\phi,\rm RH}}{\rho_{k,\rm late}}\right)^{\frac{1}{4}}+{\rm ln}\left(\frac{k}{a_{k,\rm late}H_k}\right)=0,
    \label{eq:k=aH2}
\end{align}
where $N_{\rm e}$ represents the number of e-fold during inflation and $\rho_{\phi,\rm rad}\simeq \frac{4}{3}V(\phi_{\rm end})$ is the energy density of inflaton at the end of inflation which also implies the onset of oscillation phase. $\rho_{\phi,\rm matter}$ is the energy stored in inflaton when the crossover happens from quartic to quadratic domination (at $\phi=\sqrt{2} v^\prime$) in the inflationary potential. $\rho_{\phi,\rm RH}=\frac{\pi^2}{30}g_*(T_{\rm RH})T_{\rm RH}^4$, the energy density of the Universe after the completion of reheating phase. Finally, $\rho_{k,\rm late}=\frac{\pi^2}{30}g_*(T_{k,\rm late})T_{k,\rm late}^4$ where we evaluate the $T_{k,\rm late}$ using the entropy conservation principle, $a_{k,\rm late}T_{k,\rm late}\simeq a_0T_0$. With all these inputs, the Eq.(\ref{eq:k=aH2}) relates $T_{\rm RH}$ with the quantity $N_{\rm e}$, provided the non-minimal coupling $\xi$ and $v^\prime$ are fixed. To note, $\lambda_\phi$ can be determined from the consideration of observed primordial power spectrum amplitude (using Eq.(\ref{PSpec})) which in turn specifies the Hubble scale of inflation during horizon exit of mode $k$ {\rm i.e.} $H_k$.
\begin{figure}
    \centering
 \includegraphics[scale=0.55]{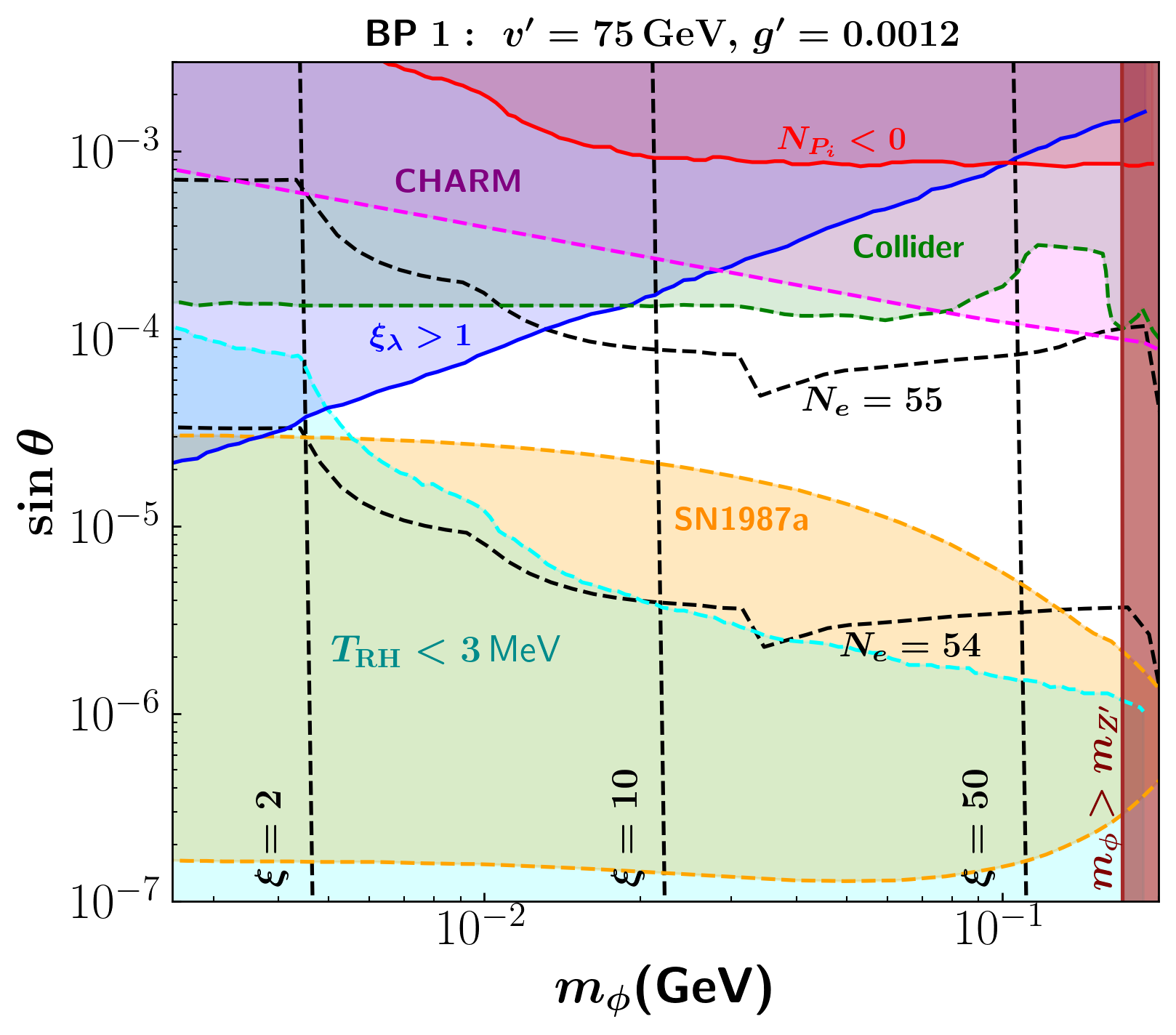}
    \caption{The correlation between inflaton sector parameters $\xi$,$N_{\rm e}$ and the scalar sector parameters is shown. Each choice of $(\xi,N_{\rm e})$ corresponds to a particular set of $(m_\phi,\sin\theta)$. Constraints from CHARM \cite{BERGSMA1985458}, SN1987a \cite{Dev:2020eam,Krnjaic:2015mbs} and Collider \cite{Balaji:2022noj,Dev:2017dui,Egana-Ugrinovic:2019wzj,Dev:2019hho} are shown in purple, orange and green colours respectively. Other forbidden regions arising from $\xi_\lambda>1$, $N_{P_i}<0$ where $i=\{{\rm rad,\,matter}\}$ and $T_{\rm RH}< 3$ MeV\, conditions are highlighted in blue, red and cyan respectively.}
    \label{fig:mphi_sinthBP1}
\end{figure}

{The Eq.(\ref{eq:k=aH2}) gives the value of $T_{\rm RH}$ as a function of $N_{\rm e}$ for a well defined inflationary potential. Now, in the particle physics model under discussion the inflaton can decay to various final state particles depending on its mass scale. We classify two cases, {\bf Case I\,:\,} $m_\phi<m_{Z^\prime}$ and {\bf Case II\,:\,} $m_\phi>m_{Z^\prime}$. In case I, the inflaton only decays to SM particles via mixing with SM Higgs with inflaton decay rate $\Gamma_{\phi_2} = \sin^2\theta \times  \Gamma_h(m_{\phi_2})$ (where $\Gamma_h(m_{\phi_2})$ is the SM Higgs decay rate when its mass is $m_{\phi_2}$), resulting the universe to enter into standard radiation domination epoch while decay channels such as $\phi\to Z^\prime \overline{l}l$ ($l$ being second and third generation SM leptons which are kinematically allowed) and $\phi\to Z^\prime Z^\prime$
remain kinematically forbidden. This choice makes the reheating temperature $T_{\rm RH}$ completely dependent on $\lambda_{\phi H}$ in case I. Once $T_{\rm RH}$ is known for a fixed $N_{\rm e}$, one can easily determine the required value of $\lambda_{\phi H}$.}

{The case II describes the dynamics of $m_\phi>m_Z^\prime$ regime. This case further includes two sub-cases, $m_{Z^\prime}<m_\phi<2m_{Z^\prime}$ and $m_\phi>2m_{Z^\prime}$. In the former case the inflaton decays via $\phi\to \overline{f}f$ (f being a representative SM particles which are
kinematically allowed) and $\phi\to Z^\prime \overline{l} l$ while for later, the additional decay mode $\phi\to Z^\prime Z^\prime$ will be present. In both the sub-cases the $T_{\rm RH}$ depends on both scalar sector parameters as well as $L_\mu-L_\tau$ gauge sector parameters.}

\begin{figure}
    \centering
    \includegraphics[height=6cm,
    width=8.2cm]{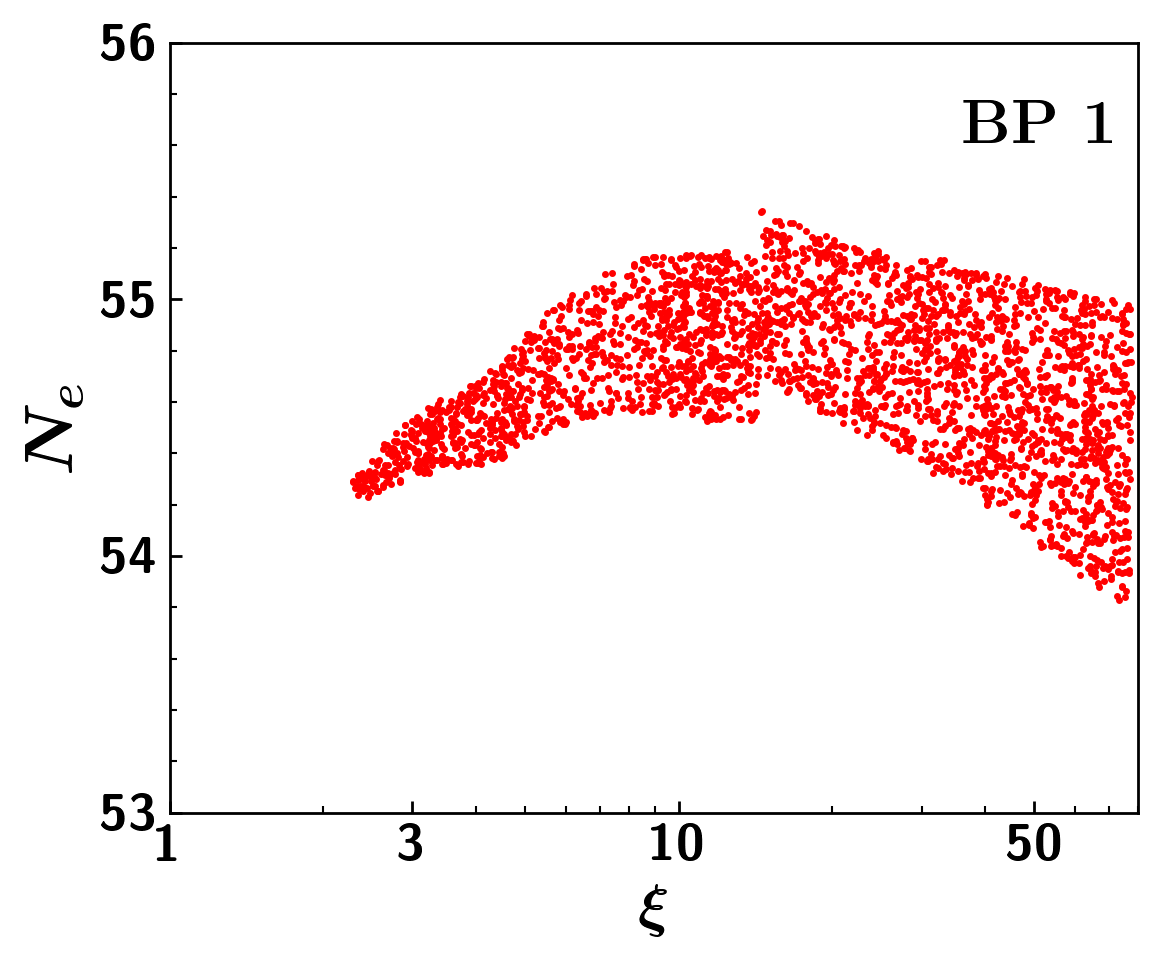}~~
    \includegraphics[height=6cm,
    width=8.2cm]{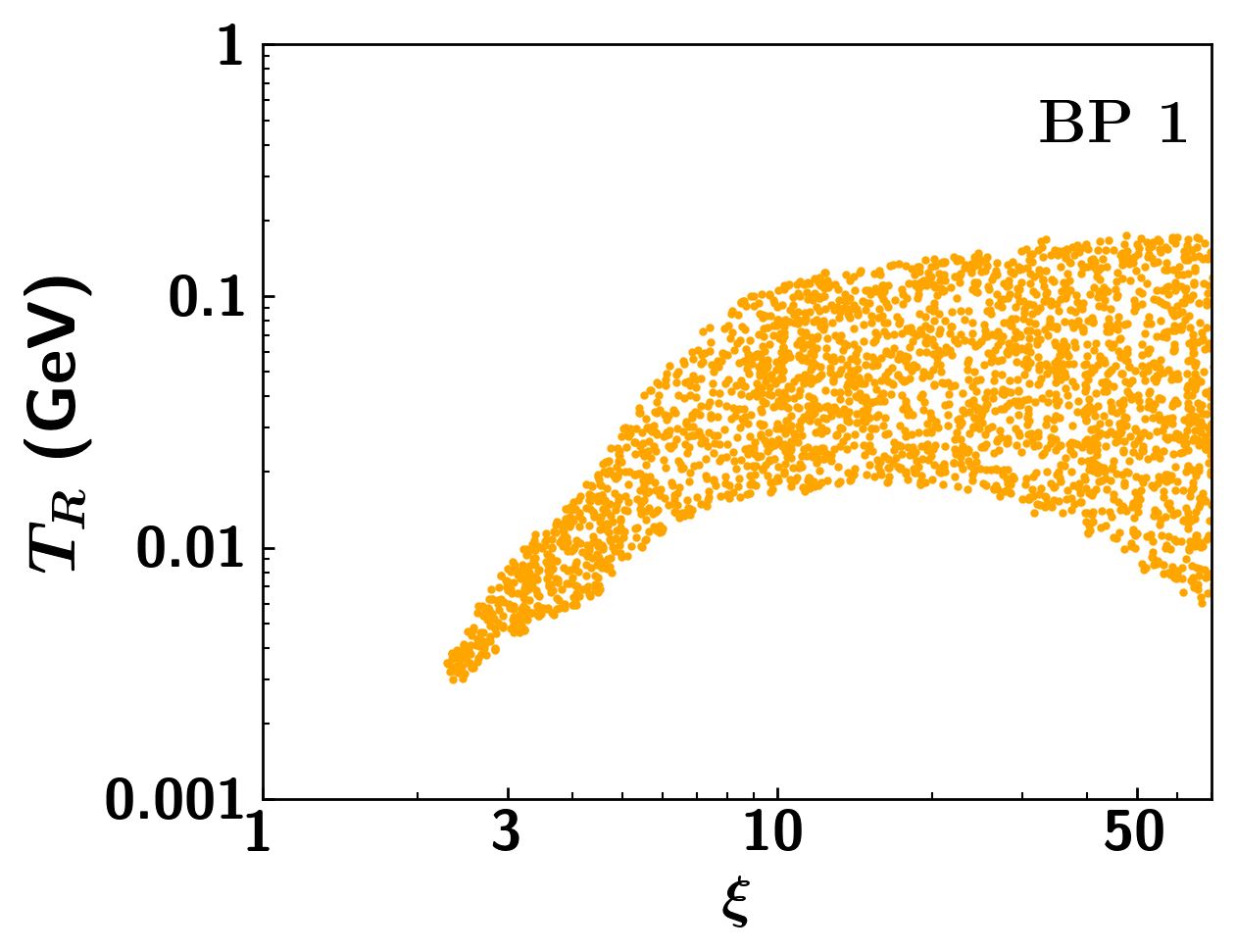}
    \caption{The allowed regions satisfying the constraints on scalar sector parameters (see Fig.\,\ref{fig:mphi_sinthBP1}) are shown in (left) $\xi-N_{\rm e}$ plane and (right) $\xi-T_{\rm RH}$ plane.}
    \label{fig:NeXi_TrhXi}
\end{figure}

\begin{figure}
\centering
    \includegraphics[scale=0.55]{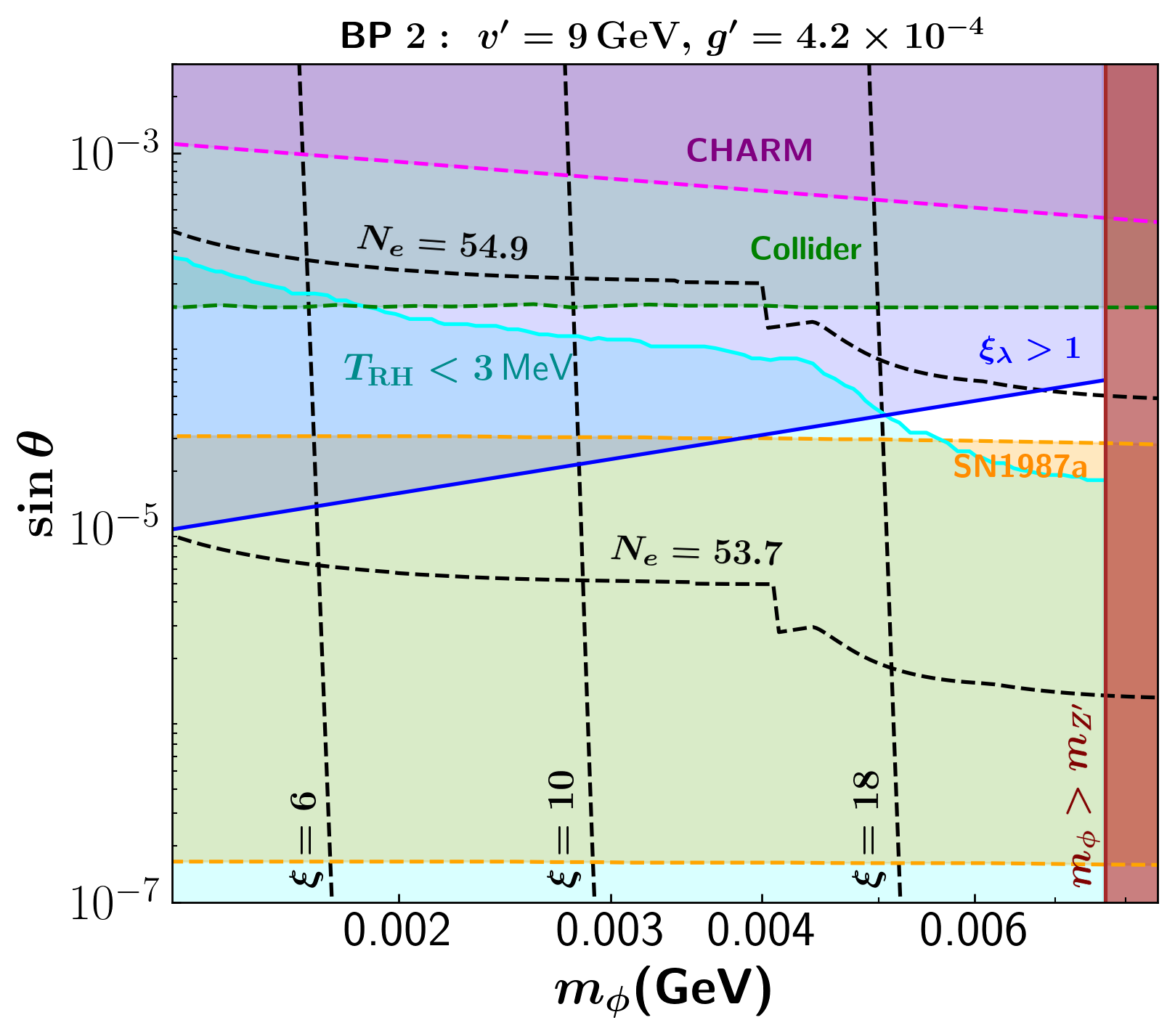}
    \caption{This plot bears same information as Fig.\,\ref{fig:mphi_sinthBP1}, but for BP\,2. Here $N_{P_i}<0$, where $i=\{\rm rad\,,matter\}$ constraint is much weaker and remains outside of the chosen ranges for $m_\phi$ and $\sin\theta$. }
    \label{fig:mphi_sinthBP2}
\end{figure}

\begin{figure}
    \centering
    \includegraphics[height=6cm,width=7.6cm]{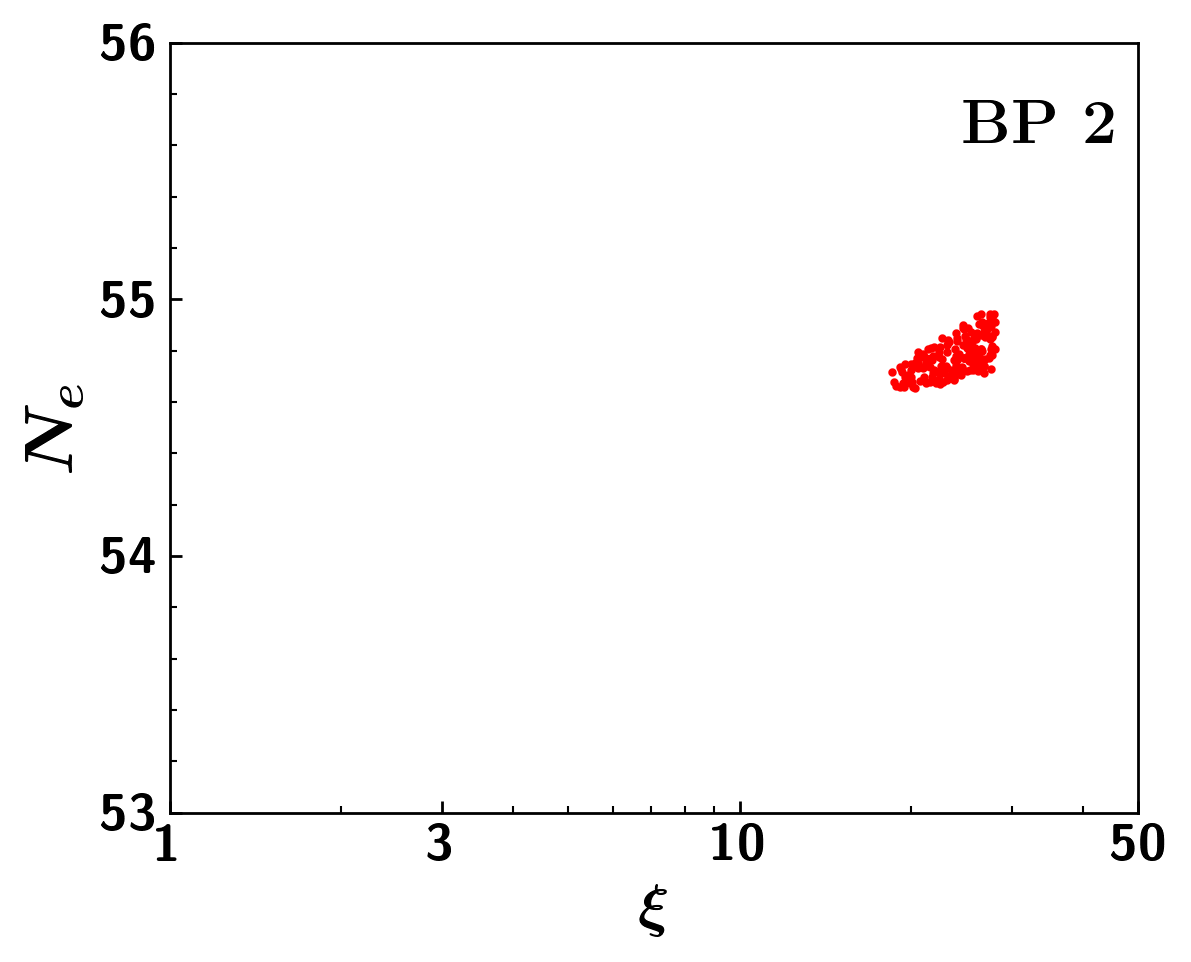}~~~
    \includegraphics[height=6cm,width=8cm]{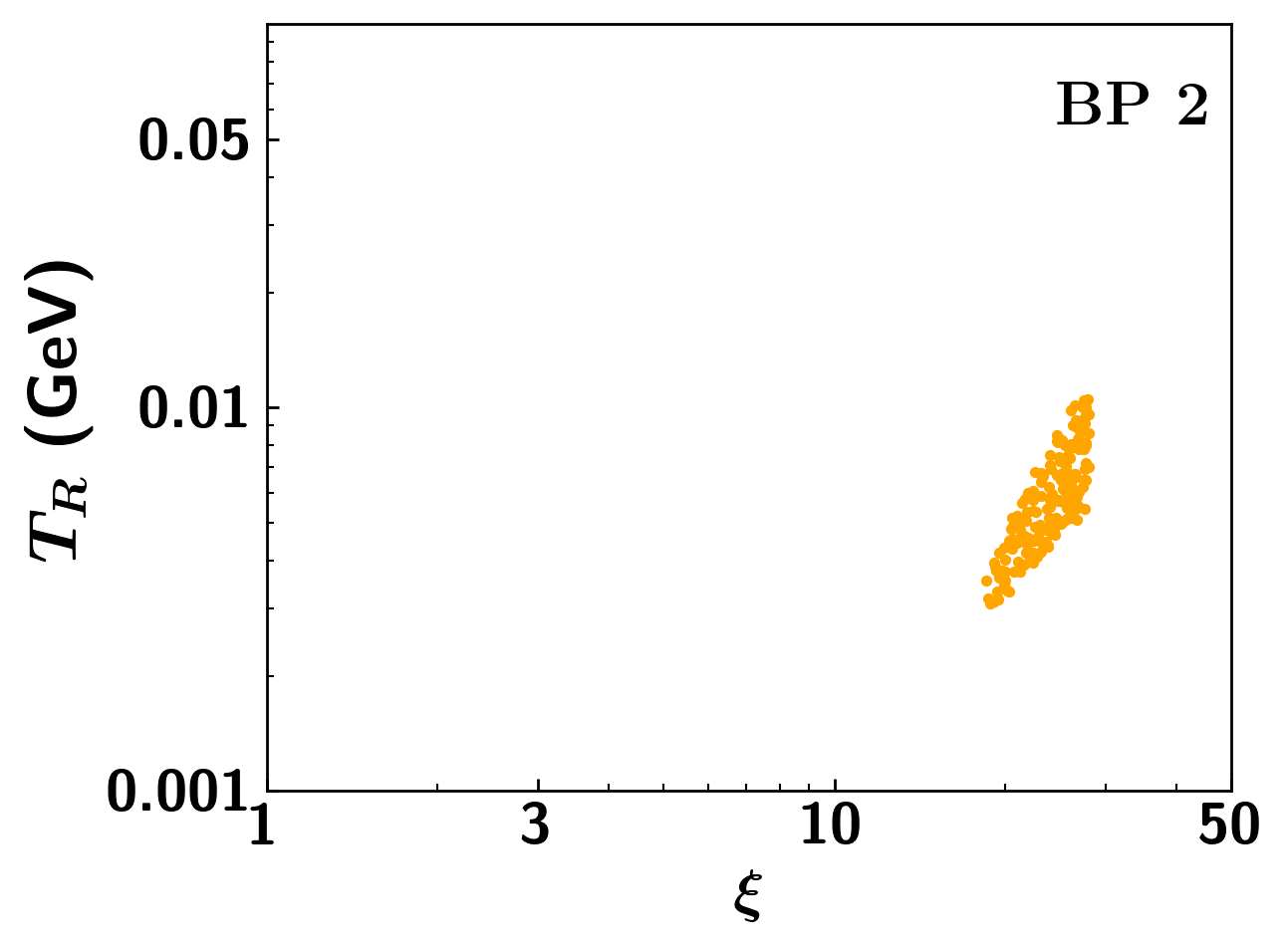}\\
    \caption{The plots of the top panel bear same information as Fig.\,\ref{fig:NeXi_TrhXi}: left and right panel respectively, however for BP\,2. 
    }
    \label{fig:othersBP2}
\end{figure}

We assume the transitions from one phase to another ({\it e.g.} radiation-matter crossover during inflaton oscillations) in this calculation as instantaneous. 
It is also possible that the gauge coupling $g^\prime$ can give rise to energy drain of inflaton into radiation via preheating mechanism well before the instantaneous perturbative reheating. We have checked numerically that even if the energy transfer is sizeable during preheating epoch, our results does not change considerably (see appendix.\,\ref{preheat} for details). In fact, the preheating solely is unable to drain the total energy density out of the inflaton field as observed from rigorous lattice simulation \cite{Podolsky:2005bw,Bernal:2018hjm,Maity:2018qhi}. This fact upholds the utmost necessity of efficient perturbative reheating via inflaton decay to set the correct initial conditions for the big bang nucleosynthesis. 



\section{Results}\label{results}
{In the preceding section we have explained how a fixed number of inflationary e-foldings can provide us an estimate of $T_{\rm RH}$. For case I, $T_{\rm RH}$ is a function of the scalar sector parameters ($m_\phi,\,\sin\theta$). On the other hand for case II, $T_{\rm RH}$ depends on both ($m_\phi,\,\sin\theta$) and $L_\mu-L_\tau$ gauge sector parameters, ($m_{Z^\prime},g^\prime$). We discuss these two cases separately.}

\subsection{Case I}
{The connection between $N_e$ and $T_{\rm RH}$ allows us to find a correlation between the set of parameters: ($\xi,\,N_{\rm e}$) and $(m_\phi,\,\sin\theta)$ as shown in Fig.\,\ref{fig:mphi_sinthBP1}. Recall that to satisfy the $(g-2)_\mu$ results, $v^\prime$ is strictly restricted to remain in the range $9{\,\rm GeV}\lesssim v^\prime \lesssim 75{\rm \,GeV}$. Note that the coupling $g^\prime$ does not have any direct impact on the predictions for inflationary observables for this case. However since we are working in the $m_\phi<m_{Z^\prime}$ regime, the chosen value of $g^\prime$ sets the maximum allowed value of $\xi$ for a particular $v^\prime$ as we will see in a while.}

{For the purpose of numerical analysis, we have considered two benchmark points, namely [BP\,1]: $v^\prime= 75\,{\rm GeV},\,g^\prime=1.2\times 10^{-3}$ and  [BP\,2]: $v^\prime=9\,{\rm GeV},\,g^\prime=4.2\times 10^{-4}$. These BPs represent the largest and smallest allowed values of $m_{Z^\prime}$ and $g^\prime$, consistent with the $(g-2)_\mu$ data as well as other relevant experimental constraints as evident from Fig.\,\ref{fig:mZP-gP plane}.}

In Fig.\,\ref{fig:mphi_sinthBP1}, we mark different constant $\xi$ and $N_{\rm e}$ lines in the $m_\phi-\sin\theta$ plane considering BP\,1. This figure reveals that a particular set of $(\xi,N_{\rm e})$ corresponds to a distinct set of ($m_\phi,\sin\theta$) values. For sub-MeV $m_\phi$, the $m_\phi-\sin\theta$ plane is restricted from different experiments, namely CHARM\,\cite{BERGSMA1985458}, SN1987a \cite{Dev:2020eam,Krnjaic:2015mbs} and colliders \cite{Balaji:2022noj,Dev:2017dui,Egana-Ugrinovic:2019wzj,Dev:2019hho}. In addition, we impose a few other important constraints related to inflaton field dynamics in the same plane which are,
\begin{itemize}
    \item $T_{\rm RH}\gtrsim 3$ MeV, as required to satisfy the BBN constraints\,\footnote{Here we have considered a sort of conservative bound on the minimum reheating temperature that does not alter the success of BBN. Lower limit on $T_{\rm RH}$ can vary depending on the decay mode (radiative or hadronic) of the parent particle as reported in \cite{Kawasaki:1999na,Kawasaki:2000en,Hasegawa:2019jsa}, however remain more or less around $\mathcal{O}(1)$\,MeV.}.
    \item $\xi\lesssim 80$, this choice of not so large non-minimal coupling $\xi$ translates into $m_\phi<m_{Z^\prime}$. This subsequently blocks the decay processes $\phi\to Z^\prime + 2 f$ and $2Z^\prime$ at tree level.
    
    \item $\xi_\lambda\lesssim1$ (with $\lambda_{H\phi}=\sqrt{4\xi_\lambda \lambda_H \lambda_\phi}$), this ensures that the scalar sector couplings $\lambda_H$ and $\lambda_{H\phi}$ do not provide significant corrections to the inflaton potential and also assures its stability \cite{Okada:2016ssd}.

    \item The respective durations during the oscillation regime of inflaton, expressed in terms of quantities $N_{\rm rad}={\rm ln}\left(\frac{a_{\phi,\rm matter}}{a_{\phi,{\rm rad}}}\right)$ and $N_{\rm matter}={\rm ln}\left(\frac{a_{\rm RH}}{a_{\phi,\rm matter}}\right)$ have to be positive in magnitude. {Violation of either or both $N_{\rm rad}>0$ and $N_{\rm matter}>0$ conditions imply that there exists no feasible solution of Eq.(\ref{eq:k=aH2}) which connects the early inflationary Universe to late time epoch through the reheating phase.}

\end{itemize}


{After imposing all these constraints in $m_\phi$\,-\,$\sin\theta$ plane, it is realised that the amount of e-foldings during inflation cannot be arbitrary for a fixed $\xi\,(\lesssim 80)$. This is clearly depicted in the left panel of Fig.\,\ref{fig:NeXi_TrhXi} where the allowed range for $N_{\rm e}$ is plotted as a function of $\xi$. In the same line, these observations also set the allowed ranges for the $T_{\rm RH}$ as a function of $\xi$ as shown in right panel of Fig.\,\ref{fig:NeXi_TrhXi}. As a numerical example, for $\xi=10$, we find $54.5\lesssim N_{\rm e}\lesssim 55.3$ and $20{\rm\,MeV}\lesssim T_{\rm RH}\lesssim 100{\rm\,MeV}$. Following a similar approach used for BP\,1, we show the favored parameter space in $m_\phi-\sin\theta$ plane in Fig.\,\ref{fig:mphi_sinthBP2} and subsequently the allowed ranges of $N_{\rm e}$ and $T_{\rm RH}$ in right and left panels of Fig.\,\ref{fig:othersBP2} considering BP 2.}

\begin{figure}
    \centering
    \includegraphics[height=6.4cm,width=8cm]{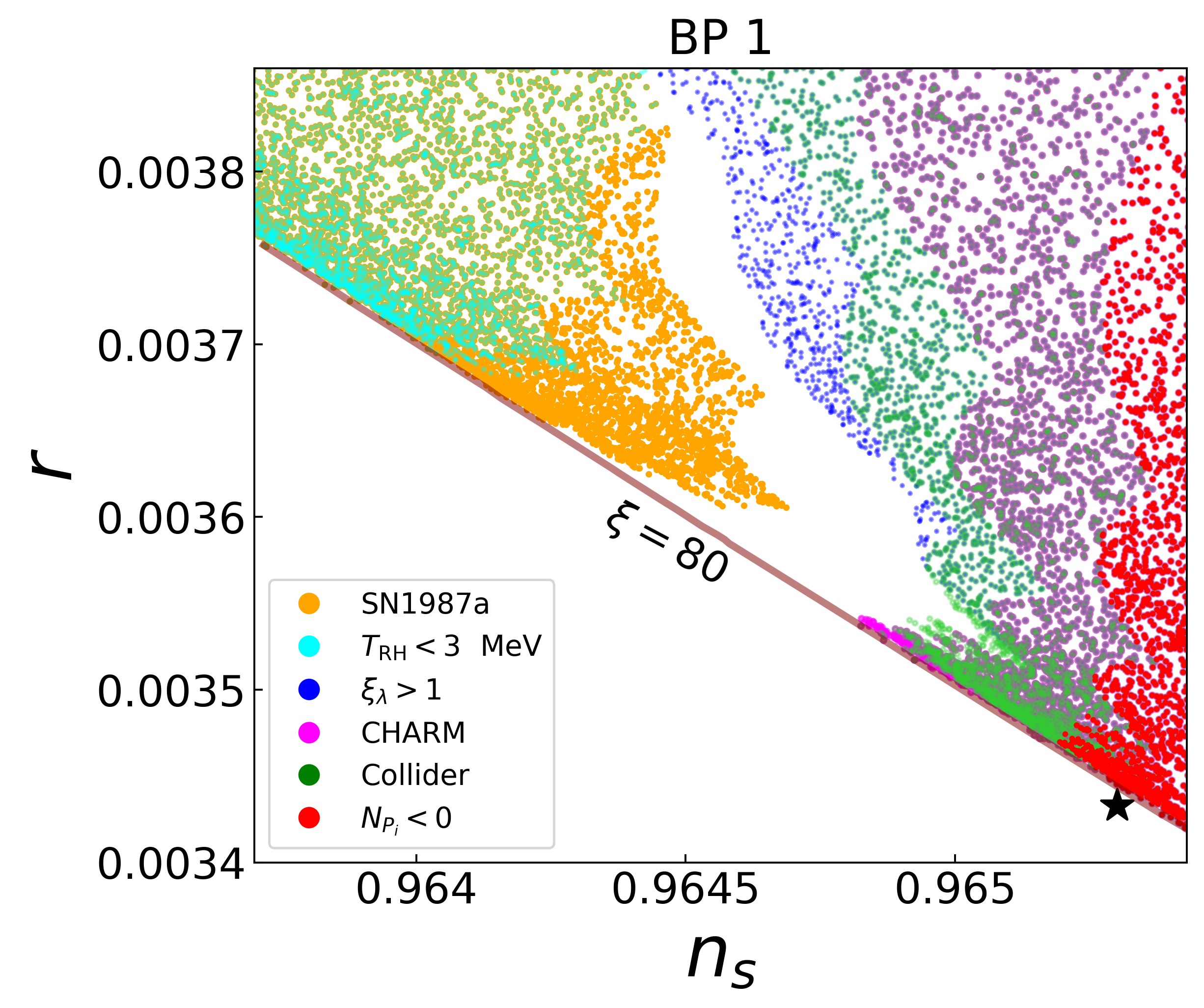}
    \includegraphics[height=6.4cm,width=8cm]{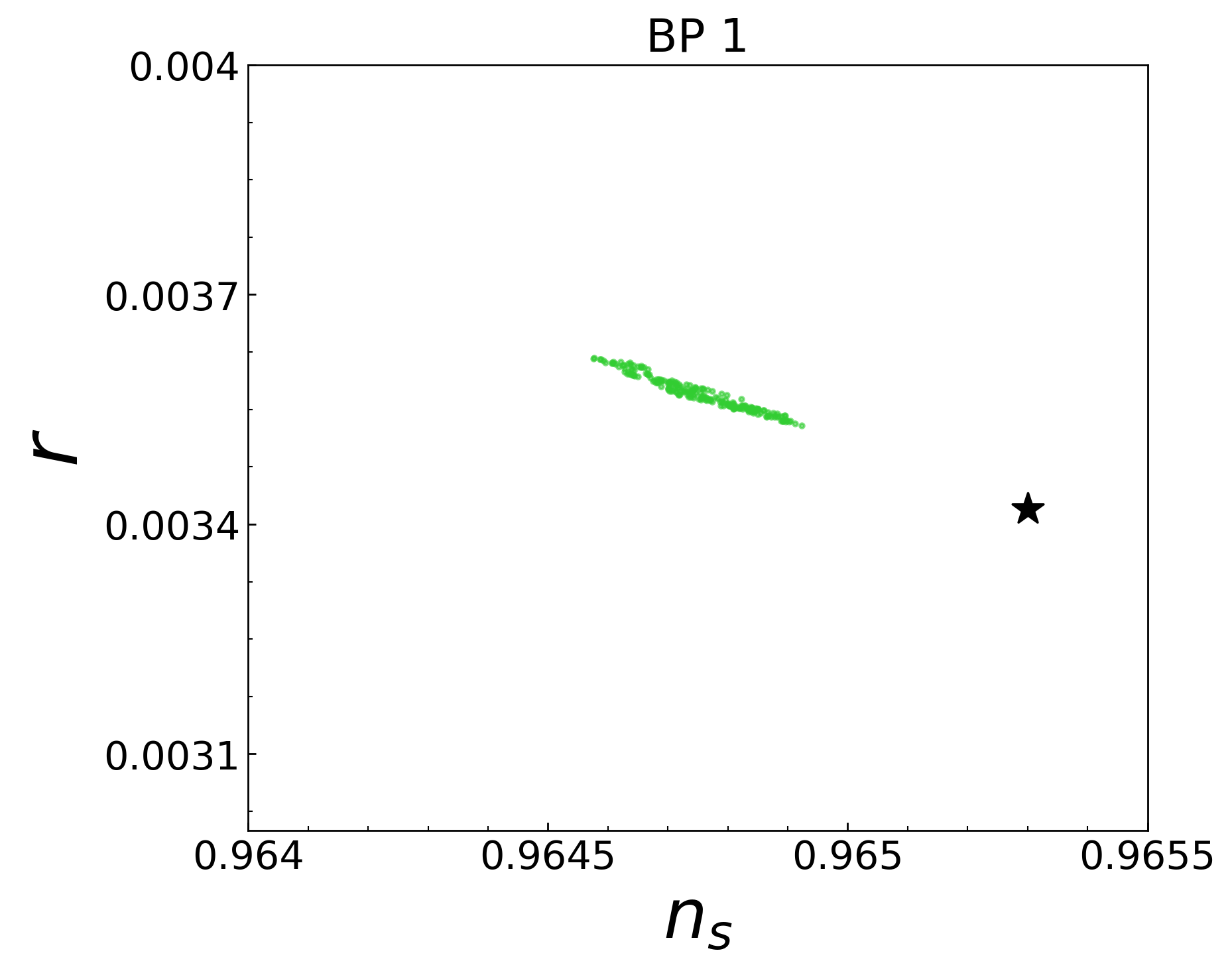}
    \caption{The predictions for $n_s$ and $r$ corresponding to BP 1 (left) and BP 2 (right). In left the allowed region (white) in the $n_s-r$ plane satisfying the constraints on the scalar sector parameters as shown in Fig.\,\ref{fig:mphi_sinthBP1}. The color codes are kept same as in Fig.\,\ref{fig:mphi_sinthBP1}. The solid brown line corresponds to $\xi=80$, which is the maximum possible value of $\xi$ in case I. In right we show the outcomes (in green) only for BP 2. The predictions for $(n_s,r)$ in case II is shown by the `$\star$' marker in both the subfigures.}
    \label{fig:Ne_TRH}
\end{figure}

We now proceed to estimate the values of spectral index and tensor to scalar ratio for both BP 1 and BP 2 following Eqs.(\ref{eq:EFold}),\,(\ref{PSpec}) and (\ref{eq:rdef}) in Case I. We present our findings in Fig.\,\ref{fig:Ne_TRH} with the left and right figures corresponding to BP 1 and BP 2 respectively. The obtained allowed ranges for $N_e$ as function of $\xi$ as presented in Fig.\,\ref{fig:NeXi_TrhXi} and Fig.\,\ref{fig:othersBP2} for BP 1 and BP 2 respectively are utilised to compute the $n_s$ and $r$. In the left, the different shaded regions (above the solid brown line) are disfavored from the relevant constraints as portrayed earlier in the $m_\phi$\,-\,$\sin\theta$ plane of Fig.\,\ref{fig:mphi_sinthBP1} with the same color codes. The region below the solid brown line corresponds  $m_\phi<m_{z^\prime}$ ($\xi>80$) which is disfavored. We end up with a small white region in the $n_s-r$ plane which corresponds to the predictions for $n_s$\,-\,$r$ for BP\,1. Next we portray the results for BP\,2 in right of Fig.\,\ref{fig:Ne_TRH}. Here we have only shown the predicted $n_s-r$ values which survive all the relevant theoretical and experimental constraints.

\subsection{Case II}\label{sec:resCII}
{In case II, the inflaton has several decay modes which are kinematically allowed. The total decay width of inflaton in this case is given by (see Appendix\,\ref{sec:DecayExp}),
\begin{align}\label{eq:GamamPhiC2}
    \Gamma_\phi=\sum_{f}\Gamma(\phi\to \overline{f}\, f) + \sum_f\Gamma(\phi\to Z^\prime\overline{l}\, l)+\Gamma(\phi\to Z^\prime Z^\prime), 
\end{align}
The Case II has two sub-cases, $m_{Z^\prime}<m_\phi< 2 m_{Z^\prime}$ and $m_\phi > 2 m_{Z^\prime}$. In the former sub-case, the $\Gamma_\phi$ includes the contribution of first two decay modes only. The decay width $\Gamma_\phi$ should be a function of $(m_\phi,\sin\theta,m_{Z^\prime},g^\prime)$ in case II for either of the benchmark points, BP\,1 and BP\,2 as defined earlier.}

{The decay modes $\phi\to Z^\prime \overline{f} f$ and $\phi\to Z^\prime Z^\prime$ do not involve any Yukawa couplings of the SM fermions at tree level. On top of that, the desired value for $g^\prime$ as fixed from $(g-2)_\mu$ data can induce rapid decay of inflaton (via $\phi\to Z^\prime \overline{f} f$ and $\phi\to Z^\prime Z^\prime$ modes) compared to the one in Case I. On the top of that a larger $\sin\theta$ (corresponding to a relatively larger $m_\phi$) boosts the inflaton decay strength further. This may cause the complete translation of energy from inflaton to radiation even before the onset of $\omega_\phi=0$ state during the inflaton oscillations. In that case, a direct transition of the Universe would occur from $\omega_\phi=1/3$ phase to $\omega=1/3$ phase with zero or minimal relevance of $\omega_\phi=0$ phase.}

{After critically investigating the evolution of each energy components (see appendix.\,\ref{sec:DecayInfC2}), we find that, the above mentioned scenario indeed happens in Case II for both BP 1 and BP 2 even with $\sin\theta=0$. Thus Eq.(\ref{eq:k=aH}) needs a little modification in order to find an estimate for $(n_s,r)$ in the present case to take into the absence of $\omega_\phi=0$ phase. The modified equation is read as,
\begin{align}
    N_{\rm e}+{\rm ln}\left(\frac{\rho_{\phi,\rm rad}}{\rho_{\phi,\rm RH}}\right)^{\frac{1}{4}}+{\rm ln}\left(\frac{\rho_{\phi,\rm RH}}{\rho_{k,\rm late}}\right)^{\frac{1}{4}}+{\rm ln}\left(\frac{k}{a_{k,\rm late}H_k}\right)=0,
    \label{eq:k=aH2C2}
\end{align}
A closer look in the last equation gives the impression that, the value of $N_e$ turns independent of $\rho_{\rm RH}$ or $T_{\rm RH}$. Therefore, the Eq.(\ref{eq:k=aH2}) is expected to provide us a uniques solution for $N_e$, regardless of the value of $g^\prime$. This feature can be understood from Fig.\,\ref{fig:evolution} as well, where it is clearly evident that absence of $\omega_\phi=0$ phase requires a fixed value of $N_e$ such that the cyan solid line matches with the blue solid line befor the horizon re-entry.}

\begin{figure}
    \centering
    \includegraphics[height=7cm]{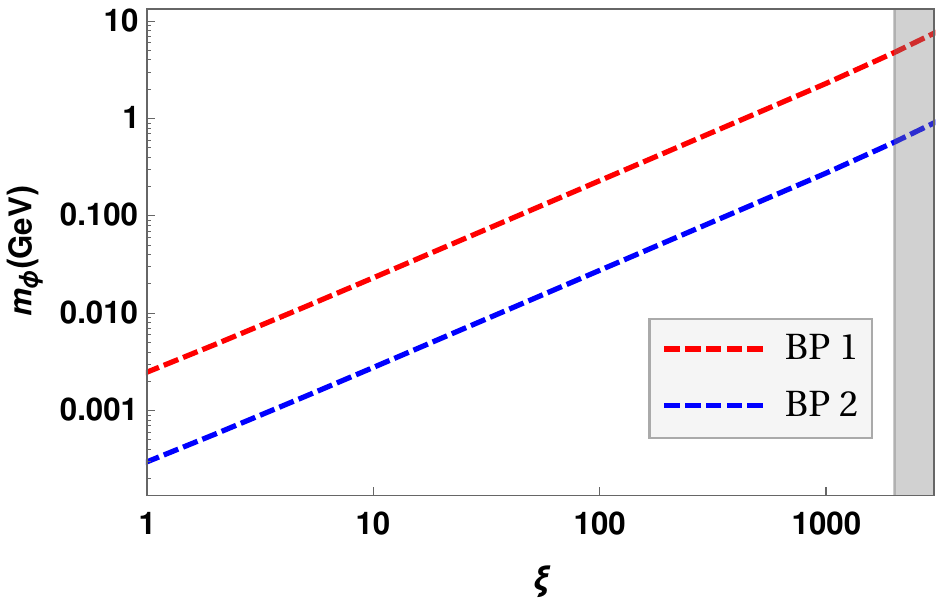}
    \caption{$m_\phi$ as a function of $\xi$ in case II considering BP 1 and BP 2. The shaded region is disfavored due to violation of unitarity bound \cite{Lebedev:2023zgw}.}
    \label{fig:mphiXi}
\end{figure}

{The non-minimal coupling $\xi$ poses a correlation with the mass of the inflaton due to the involvement of $\lambda_\phi$ in both the quantities (see Fig.\,\ref{fig:mphiXi}). Note that in a generic single field inflation model with non-minimal coupling to inflaton, the value of $\xi$ ($\lesssim 2000$) is strictly restricted by the unitarity bound \cite{Lebedev:2023zgw}. This further imposes an upper-bound on the mass of the inflaton in the present framework as given by,
\begin{equation}
  m_\phi \lesssim
    \begin{cases}
      4.86{\rm~GeV} & \text{for BP 1}\\
      0.56{\rm~GeV} & \text{for BP 2}
    \end{cases}       
\end{equation}
Within this mass range, the inflaton decay width for Case II is always dominated by the decay modes having $Z^\prime$ in the final states.}

\begin{figure}
    \centering
    \includegraphics[height=6cm]{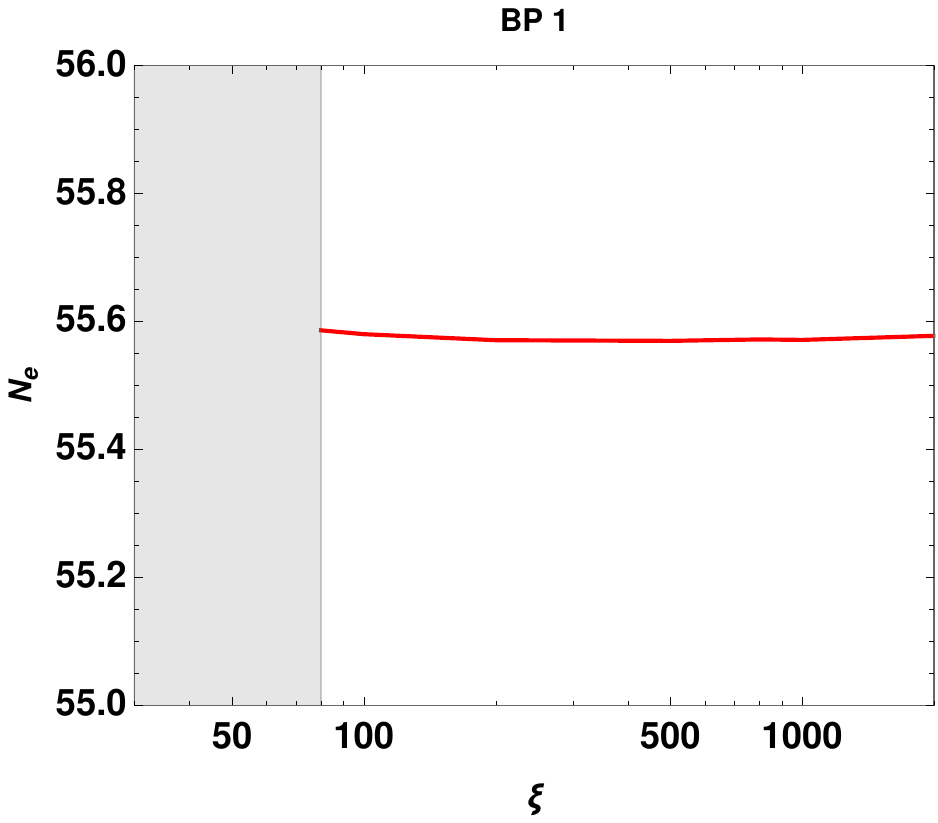}
    \includegraphics[height=6cm]{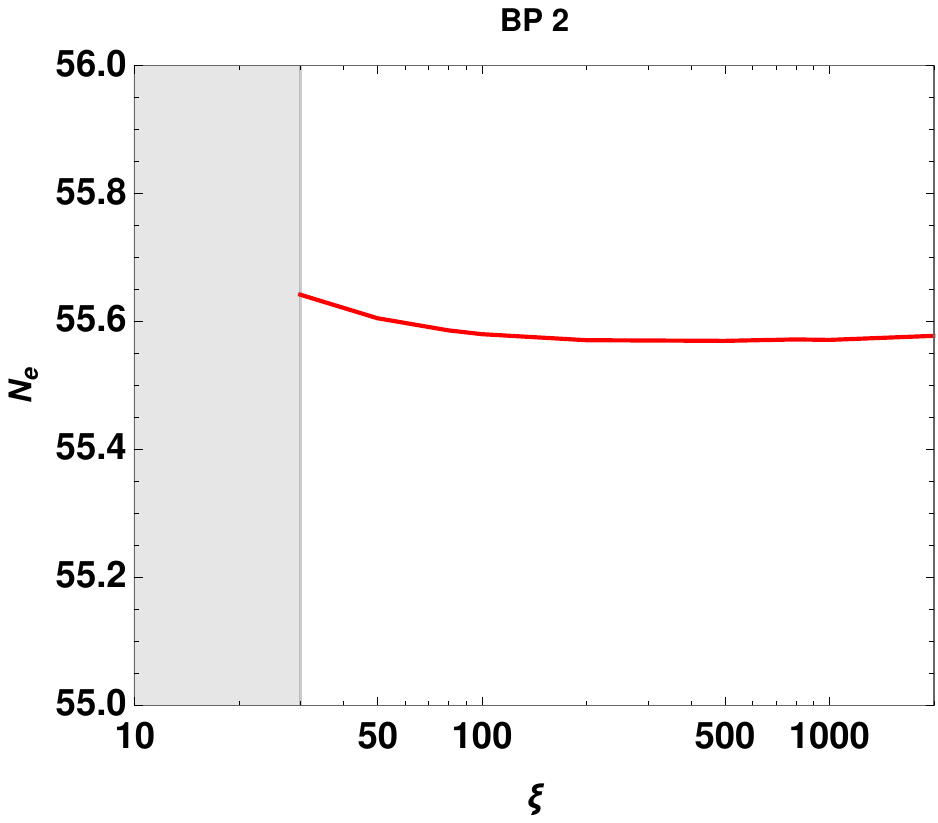}\\
    \includegraphics[height=6cm]{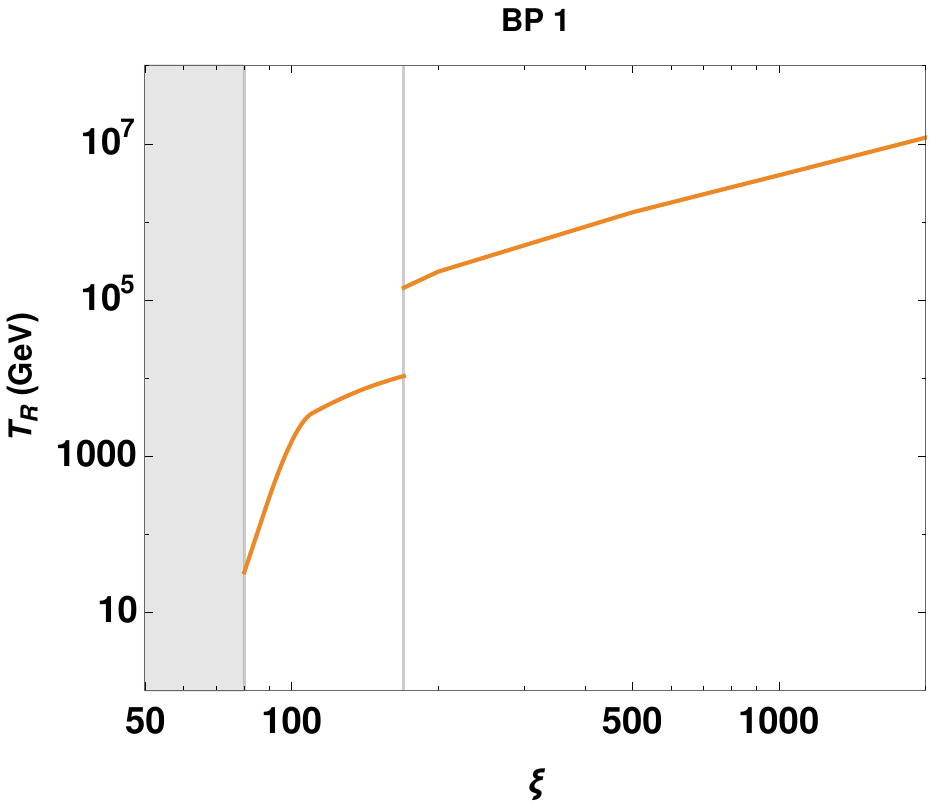}
    \includegraphics[height=6cm]{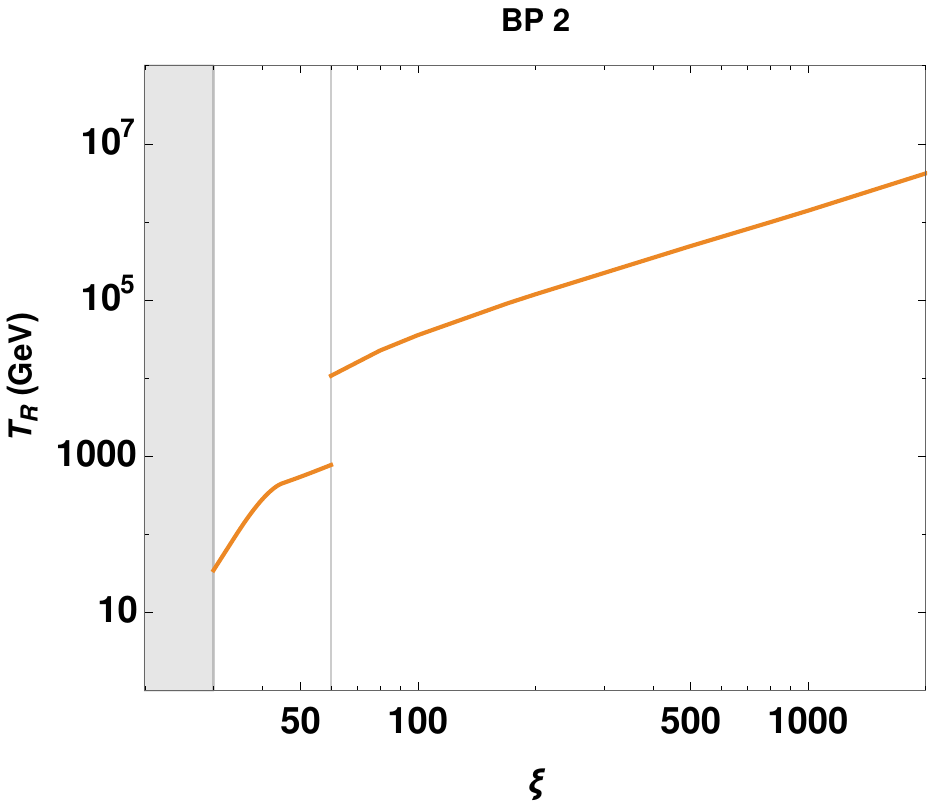}\\
    \caption{Predictions of number of e-folds during inflation (top) and $T_R$ during inflation (bottom) are shown as a function of $\xi$ considering $m_\phi>m_{Z^\prime}$. In each of the sub-figures, the shaded region corresponds to the region pertaining to Case I.}
    \label{fig:NeTRBC2}
\end{figure}

{In Fig.\,\ref{fig:NeTRBC2}, we have highlighted the predictions of $N_e$ and $T_R$ for both BP 1 and BP 2 in case II. As anticipated earlier, $N_e$ shows almost no-variation with varying $\xi$. In fact the requisite value of $N_e$ remains almost same for the BPs as it does not depend on the reheating history for Case II following Eq.(\ref{eq:k=aH2C2}). This will be further clear from the semi analytical computation of inflationary observables provided in sec.\,\ref{sec:semiANA} (see Eq.(\ref{eq:nsC2})). The constant nature of $N_e$ upon the variation of $\xi$ and $T_{\rm RH}$ for both the BPs results in an unique prediction for $(n_s,r)$ as given by,
\begin{align}
    n_s=0.9653,~r=0.00342
\end{align}
This prediction has been highlighted in the form of a black dot in left and right panels of Fig.\,\ref{fig:Ne_TRH}.}

\begin{figure}
    \centering
    \includegraphics[width=0.8\textwidth]{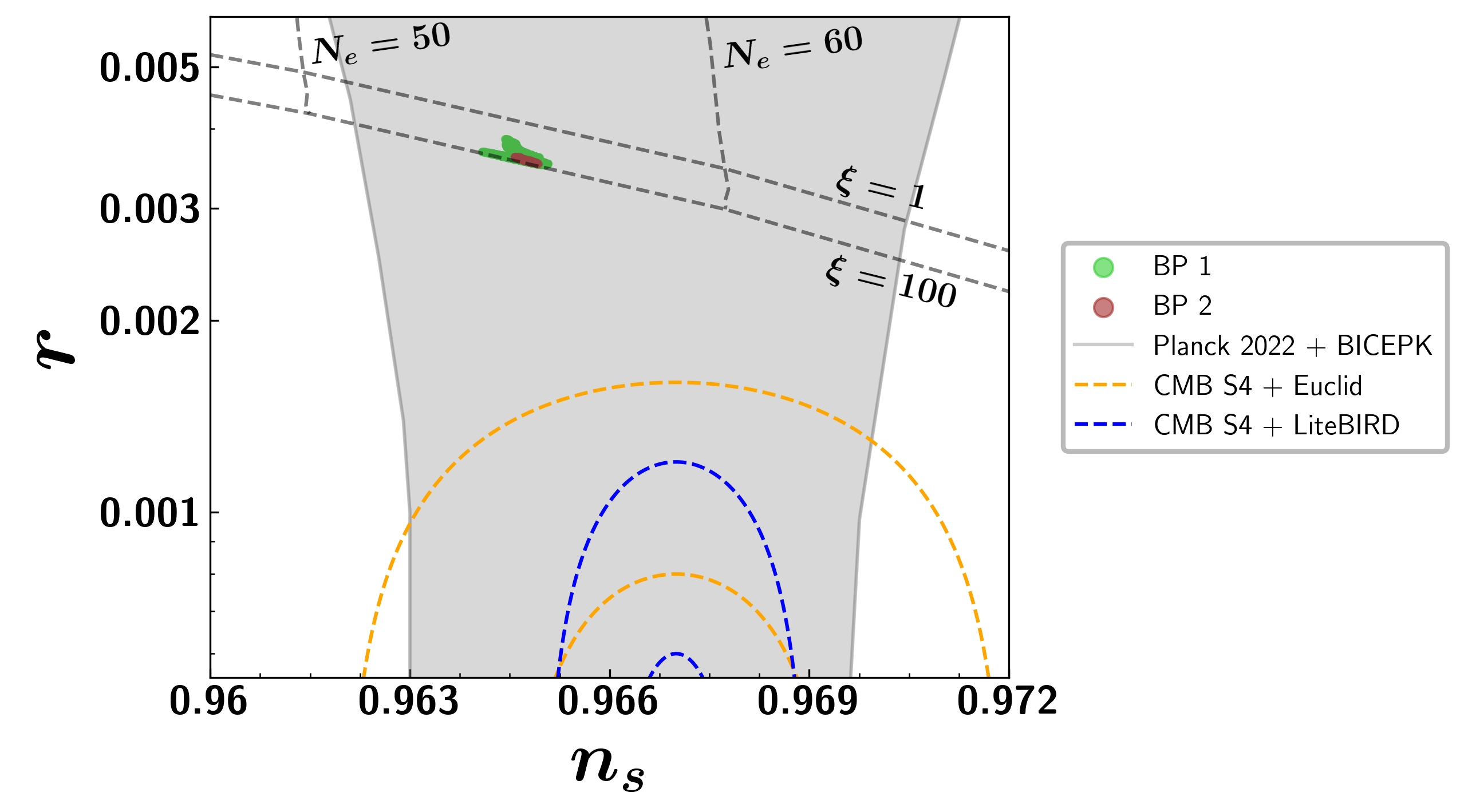}
    \caption{$n_s-r$ predictions for BP\,1 (green) and BP\,2 (brown) in Fig.\,\ref{fig:mZP-gP plane}. The Planck+BICEPK \cite{Planck:2018jri,BicepKeck:2021ybl} 1-$\sigma$ allowed region is indicated by gray. Blue and orange contours correspond to 1-$\sigma$ (dashed) and 2-$\sigma$ (dotted) future sensitivites of CMB\,S4 \cite{CMB-S4:2020lpa}+Euclid \cite{laureijs2011euclid} and CMB\,S4+LiteBIRD \cite{LiteBIRD:2022cnt} respectively assuming mean value of $n_s=0.967$ and $r=0$. The gray dashed and dotted lines are the estimates of $n_s-r$ corresponding to a generic non-minimal quartic inflation model.}
    \label{fig:nsr}
\end{figure}

Finally, we highlight our findings of $n_s-r$ {after combining Case I and Case II} against the existing Planck data as well as future observations from CMB\,S4 \cite{CMB-S4:2020lpa} and LiteBIRD \cite{LiteBIRD:2022cnt} {\it etc.} experiments in Fig.\,\ref{fig:nsr}. For comparison purpose, we also provide the $n_s-r$ values corresponding to a generic non-minimal quartic inflation for $\xi=1$ and 100 as indicated by dotted lines. As observed, the $n_s-r$ predictions for both BP\,1 and BP\,2 are much more restricted than the one in non-minimal quartic inflation due to the involvement of $(g-2)_\mu$ data in combination with all other relevant theoretical and experimental constraints in the $m_{Z^\prime}-g^\prime$ and $m_\phi-\sin\theta$ planes. It is also noticed that $n_s-r$ values corresponding to BP\,2 is a subset of the one corresponding to BP\,1. We have also confirmed that for any random benchmark point in the $m_{Z^\prime}-g^\prime$ plane (Fig.\,\ref{fig:mZP-gP plane}), the obtained estimates of $n_s-r$ always remain inside the predicted region for BP\,1 and therefore we conclude that $(g-2)_\mu$ data in the minimal $L_\mu-L_\tau$ model can at most allow $0.964\lesssim n_s\lesssim 0.965$ with $0.0035\lesssim r\lesssim 0.0039$. Future CMB experiments like CMB\,S4 \cite{CMB-S4:2020lpa} and LiteBIRD \cite{LiteBIRD:2022cnt} {\it etc.}
will decide the fate of viability of cosmic inflation in the minimal $L_\mu-L_\tau$ model accommodating the $(g-2)_\mu$ data.
\section{Semi-analytic calculation}\label{sec:semiANA}
{In this section, we provide a semi-analytic calculation that helps in estimating the inflationary observables in a  relatively simple manner. From Eq.(\ref{eq:fieldTrans}), a general analytical form of $\chi$ as a function of $\phi$ is difficult to obtain. Instead, considering different regimes of $\phi$ field value, we find the analytical form for $\chi$ and subsequently the inflationary potential.

\noindent $\bullet$ $\phi<\frac{M_P}{\sqrt{6}\xi},\,$ in this limit we find approximately,
\begin{align}
\frac{d\chi}{d\phi}\approx 1,\,\,\Omega^2\sim 1\, \text{and}\,\, \chi\simeq \phi.
\end{align}
This implies,
\begin{align}
V(\chi)\simeq \frac{\lambda_\phi}{4}\chi^4.
\end{align}
 
\noindent $\bullet$ $\frac{M_P}{\sqrt{6}\xi}<\phi<\frac{M_P}{\xi}$,\, in this regime we find approximately,
\begin{align}
\frac{d\chi}{d\phi}\approx \frac{\sqrt{6}\xi\phi}{M_P},\,\,\Omega^2\sim 1\, \text{and}\,\, \chi\simeq \frac{\sqrt{6}\xi\phi^2}{2M_P}.
\end{align}
This potential for $\chi$ is given by,
\begin{align}
V(\chi)\simeq \frac{\lambda_\phi}{6}\frac{M_P^2
\chi^2}{\xi^2}.
\end{align}

\noindent $\bullet$ $\phi>\frac{M_P}{\xi}$,\, in this case we find approximately,
\begin{align}
\frac{d\chi}{d\phi}\approx \frac{\sqrt{6}M_P}{\phi},\,\,\Omega^2\sim \frac{\xi\phi^2}{M_P^2}\, \text{and}\,\, \chi\simeq \sqrt{6}M_P\text{ln}
\left(\frac{\sqrt{\xi}\phi}{M_P}\right).
\end{align}
This inflationary potential is then given by,
\begin{align}
V(\chi)\simeq \frac{\lambda_\phi}{4}\frac{M_P^4}{\xi^2}\left[1-\text{exp}\left(-\frac{2\chi}{\sqrt{6}M_P}\right)\right]^2.
\end{align}

The third case can provides a flat potential at large inflaton field values and can succesfully drive the cosmic inflation at early Universe. We define $\Lambda^4=\frac{\lambda_\phi M_P^4}{4 \xi^2}$ and evaluate the approximate analytial form of all the quantities, relevant to describe cosmic inflation.
We first compute the spectral index ($n_s$) under the slow roll approximation,
\begin{align}
&n_s=1-\frac{8\left(e^{\sqrt{\frac{2}{3}}\frac{\chi_k}{M_P}}+1\right)}{3\left(e^{\sqrt{\frac{2}{3}}\frac{\chi_k}{M_P}}-1\right)^2},
\end{align}
where $\chi_k$ is the inflaton field value at the horizon exit. We now express $\chi_k$ as a function of $n_s$.
\begin{align}\label{eq:phikdef}
~\chi_k=\sqrt{\frac{3}{2}}\,M_P\,{\rm ln}\left(1+\Delta(n_s)\right)\simeq 0.94 M_P,
\end{align}
where $\Delta(n_s)=\frac{4}{3}\frac{1+\sqrt{1+3(1-n_s)}}{(1-n_s)}$. Next, we estimate the inflaton field value at the end of inflationary epoch by equating one of the slow roll parameters (max[$\epsilon,\eta$]) to unity as given by,
\begin{align}
\chi_{\rm end }=\sqrt{\frac{3}{2}}\,M_P\,{\rm ln}\left(\frac{2}{\sqrt{3}}+1\right).\label{eq:phiEND}    
\end{align}
Using Eq.(\ref{eq:phikdef}) and Eq.(\ref{eq:phiEND}), we 
derive the forms of $r$ and $N_e$ (originally defined in Eq.(\ref{eq:rdef}) and Eq.(\ref{eq:Ne}) respectively)
\begin{align}
  r&=\frac{64}{3\left(e^{\sqrt{\frac{2}{3}}\frac{\chi_k}{M_P}}-1\right)^2}\,=\,\frac{48(1-n_s)^2}{\left[1+\sqrt{1+3(1-n_s)}\right]^2},\label{eq:r}\\
  N_k&=\frac{3}{4}\left[e^{\sqrt{\frac{2}{3}}\frac{\chi_k}{M_P}}-e^{\sqrt{\frac{2}{3}}\frac{\chi_{\rm end}}{M_P}}-\sqrt{\frac{2}{3}}\frac{(\chi_k-\chi_{\rm end})}{M_P}.\right]\label{eq:Nk}
\end{align}
The scalar potential at the end of inflation and the scalar perturbation spectrum are obtained as,
\begin{align}
& V(\chi_{\rm end})= \Lambda^4\left(\frac{2}{2+\sqrt{3}}\right)^{2},\label{eq:Vend}\\
 & A_s=\frac{2 V(\phi_k)}{3\pi^2r M_P^4}.
\end{align}
The observed value of $A_s^{\rm obs}=2.2\times 10^{-9}$ \cite{Planck:2018jri} precisely fixes one of the model parameters $\Lambda$
as,
\begin{align}\label{eq:Lambdadef}
    \Lambda=& M_P\left(\frac{3\pi^2 r A_s^{\rm obs}}{2}\right)^{1/4} \times\left[\frac{3+\sqrt{4-3n_s}}{4}\right]^{1/2}.
\end{align}
Note that, all of the above quantities ($r,N_k,V_{\rm end} {~\rm and~} \Lambda$) have been expressed exclusively as a function of spectral index $n_s$.

 Equipped with these, let us try to simplify Eq.(\ref{eq:k=aH2}) that unifies the different primordial events of the Universe starting from inflation to the present Universe through reheating phase in a single thread {corresponding to case I}. We write,
\begin{align}\label{eq:MainInf}
N_e+N_{\rm rad}+N_{\rm matter}+N_k^{\rm late}+{\rm ln}\left(\frac{k}{a_{k,\rm late}H_k}\right)=0,
\end{align} 
where \footnote{At the end of inflation, $\chi$ turns sub-Planckian (see Eq.(\ref{eq:phikdef})) and hence in the post-inflationary period $\phi\equiv \chi$.},
 \begin{align}
 N_{\rm rad}={\rm log}\left(\frac{\rho_{\phi,\,\rm rad}}{\rho_{\phi,\,\rm matter}}\right)^{\frac{1}{4}},\,N_{\rm matter}={\rm log}\left(\frac{\rho_{\phi,\,\rm matter}}{\rho_{\phi,\,\rm RH}}\right)^{\frac{1}{3}},\,N^k_{\rm late}={\rm log}\left(\frac{\rho_{\phi,\,\rm RH}}{\rho_{k,\,\rm late }}\right)^{\frac{1}{4}},
 \end{align}
and,
\begin{align}
& \rho_{\phi,\rm rad}\simeq \frac{4}{3}V(\phi_{\rm end}),\, \rho_{\phi,\rm matter}=\frac{2\Lambda^4\xi^2 v_\phi^4}{M_P^4},\,\rho_{\phi,\rm RH}=\frac{\pi^2}{30}g_*(T_{\rm RH})T_{\rm RH}^4,\nonumber \\
& \rho_{k,\rm late}=\frac{\pi^2}{30}g_*(T_{k,\rm late})T_{k,\rm late}^4,\,\,\text{with}\,\,T_{k,\rm late}=\frac{a_0T_0}{a_{k,\,\rm late}},H_k=\frac{\pi M_P\sqrt{r A_s^{\rm obs}}}{\sqrt{2}}\,.
\end{align}
After replacing all the defined quantities in Eqs.(\ref{eq:phikdef})-(\ref{eq:Lambdadef}) in Eq.(\ref{eq:MainInf}), we obtain,
\begin{align}\label{eq:anaL}
&60.0327-\frac{\sqrt{4-3n_s}+1}{1-n_s}+\frac{3}{4} \log \left(\frac{7-3 n_s+4 \sqrt{4-3 n_s}}{3 (1-n_s)}\right)\nonumber\\
&~~~-\frac{1}{3} \log \left(\frac{\left(\sqrt{4-3 n_s}+1\right) \left(\sqrt{4-3 n_s}+3\right)^2}{1-n_s}\right)-\frac{1}{3} \log \left(\frac{\sqrt{\xi} v^\prime}{0.173 \sqrt{\Gamma_\phi M_P}}\right)=0,
\end{align}
where we have used the relation $T_{\rm RH}\sim \frac{1.73}{g_*(T_{\rm RH})}\sqrt{\Gamma_\phi M_P}$ and $g_*\simeq 10$ assuming $\mathcal{O}(10)$ MeV reheating temperature. Note that the inflaton decay width is proportional to $m_\phi$ which has te be expressed as,
\begin{align}
m_\phi=\sqrt{2\lambda_\phi} v^\prime=\frac{2\sqrt{2}\xi\Lambda^2}{M_P^2}v^\prime,
\end{align}
while estimating $\Gamma_\phi$ (see sec.\,\ref{sec:DecayExp}).

Eq.(\ref{eq:anaL}) cleary depicts the connection between the value of spectral index and the decay width of inflaton which is function of the $L_\mu-L_\tau$ gauge sector parameters, $m_\phi,\sin\theta,g^\prime$. The Eq.(\ref{eq:anaL}) also provides a simple way to determine the $n_s$ value for a particular set of $(\xi,v^\prime,\Gamma_\phi)$. 
Once $n_s$ is known all the other observables namely $r$, $N_e$ {\it etc.} can be easily computed from the analytical expressions defined earlier. For example, in {case I}, considering BP 1 and $m_\phi>m_Z^\prime$ case, we set $\xi=10$ and $\sin\theta=5\times 10^{-5}$. We obtain the following estimates,
\begin{align}
m_\phi=\sqrt{2\lambda_\phi} v^\prime\simeq22.9\,{\rm MeV},\,\, N_e=54.78,\,\,T_{\rm RH}=7.8\,{\rm MeV}\,\,n_s=0.9648,\,\,r=0.00352\,,
\end{align} 
which are more or less consistent with our earlier results obtained numerically.


{In case II, the $\omega_\phi=0$ phase is absent and the equation that unifies the different primordial events of the Universe starting from inflation to the present Universe through reheating phase is,
\begin{align}
N_e+\frac{1}{4}{\rm ln}\left(\frac{\rho_{\phi,\,\text{rad}}}{\rho_{k\,{\rm late}}}\right)+\text{ln}\left(\frac{k}{a_kH_k}\right)=0.
\end{align}
This translates into the following equation,
\begin{align}\label{eq:nsC2}
    &58.93-\frac{\sqrt{4-3n_s}+1}{1-n_s}-\frac{1}{2} \log \left(\frac{\left(\sqrt{4-3 n_s}+1\right) \left(\sqrt{4-3 n_s}+3\right)}{1-n_s}\right)\nonumber\\
    &~~~~~~~~~~~~~~~~~~~~~+\frac{3}{4} \log \left(\frac{-3 n_s+4 \sqrt{4-3 n_s}+7}{3-3 n_s}\right)=0
\end{align}
Clearly, in case II, the value of $n_s$ (and $r$) do not depend on the parameters that determine the reheating temperature of the Universe unlike in case I. Solution of Eq.(\ref{eq:nsC2}) yields,
\begin{align}
n_s=0.9655, r=0.0034
\end{align}
which is more or less consistent with our earlier numerical findings.}

The dependence of the inflationary observables on $L_\mu-L_\tau$ gauge parameters, also constrained by the explanation of $(g-2)_\mu$ anomaly makes the outcomes of present inflationary scenario different from the usual Higgs or Higgs-like inflation. In the conventional Higgs inflation, the number of e-folds during inflation can vary in the range (not very strictly) $N_e=50-60$ and it is not constrained with the low energy data {\it e.g. $(g-2)_\mu$} or gauge sector parameters in contrast to the present scenario. Additionally $\xi$ can take any values in an unconstrained manner. In case of Higgs or Higgs-like inflation, the value of $n_s$ can vary in the range $0.9615-0.9675$ and $0.003<r<0.005$ corresponding to $50<N_e<60,\,\text{~and~} \xi=1-100$ (dashed lines in (Fig.\,\ref{fig:nsr})) whereas we find a very predictive and small allowed region in the number of e-folds during inflation and subsequently on $n_s-r$ plane after taking into account the $(g-2)_\mu$ data in the gauged $U(1)_{L_{\mu-\tau}}$ model. Similar conculsion can be drawn about the estimate of tensor to scalar ratio ($r$) as well. This stems from that fact that the reheating temeperature and non-minimal coupling parameter $xi$ are strongly restricted by different particle physics experiments like CHARM, SN1987a {\it etc.}. Our analysis reveals that accomodating cosmic inflation in the $(g-2)_\mu$ satisfying parameter space of the $U(1)_{L_\mu-L_\tau}$ gauged model is indeed possible with very predictive outcomes in the $n_s-r$ plane. In a different perspective, our study also indicates a possible cosmological probe of $(g-2)_\mu$ parameter space in the $U(1)_{L_\mu-L_\tau}$ gauged model thus complementing the ongoing and future particle physics experiments.}

\section{Conclusion and Discussions}\label{conclusion}
In this work, we analyse the compatibility of minimal gauged $L_\mu-L_\tau$ model in accommodating cosmic inflation, consistent with the $(g-2)_\mu$ data. We identify the additional SM gauge singlet scalar as the inflaton which is non-minimally coupled with the gravity. We observe that satisfaction of $(g-2)_\mu$, consistent with the other existing experimental constraints set the preferred range for the vev of the additional scalar. The non-zero vev of the scalar has  two fold non-trivial roles in the post inflationary evolution of the Universe. First of all, $v^\prime$ determines the shape of the inflaton potential at low field values, relevant for inflaton oscillation. {Secondly, $v^\prime$ enters into the mass parameter of inflaton that serves crucial role in controlling the reheating temperature. Two possible case studies have been performed. In Case I, relevant to $m_\phi < m_Z^\prime$, the reheating temperature is determined by $L_\mu-L_\tau$ scalar - SM Higgs mixing along with the inflaton mass. On the other hand in case II, pertaining to $m_\phi>m_{Z^\prime}$ regime, there exist additional decay channels of the inflaton field into final state particles involving $L_\mu-L_\tau$ gauge boson. {It finally turns out that the Case II provides an unique predictions for $(n_s,r)$, ensuring a consistent cosmological history starting from inflation through the reheating phase to late time epoch}.

{We utilise the principle that a viable cosmological history of the Universe uniquely predicts the reheating temperature of the Universe (subsequently $\sin\theta$ corresponding to a particular $m_\phi$) for a given inflationary number of e-folds.} This after taking into account all possible phenomenological constraints in the $m_\phi-\sin\theta$ plane significantly constrains the allowed reheating temperature of the Universe and also the number of inflationary e-folds. Using the bound on inflationary number of e-folds we further compute the spectral index and tensor to scalar ratio. We find satisfaction of $(g-2)_\mu$ data allows $n_s$ and $r$ to remain in a very narrow region compared to the ones in a generic non-minimal quartic inflationary set up and is refutable by future CMB experiments with improved sensitivities.

{One may wonder about the scope of yielding the observed amount of baryon asymmetry, given the fact that the present framework predicts reheat temperature to remain substantially below the electroweak phase transition temperature. Indeed the conventional mechanism of producing baryogenesis via leptogenesis \cite{Buchmuller:2004nz} from heavy sterile neutrino decay does not work here. Recently a new paradigm of low-scale baryogenesis has been proposed \cite{Elor:2018twp,Elor:2020tkc,Alonso-Alvarez:2021qfd}, known as {\it mesogenes}is at $\mathcal{O}(10)$ MeV temperatures. In this scenario, late time out of equilibrium decay of a BSM singlet scalar produces SM quarks that hadronize at low temperatures to form neutral and charged mesons. The produced mesons ($B$\,\cite{Elor:2018twp,Alonso-Alvarez:2021qfd,Elahi:2021jia} and $D$,\,\cite{Elor:2020tkc} mesons in SM) further decay to dark and SM baryons through various CP violating processes, thus can potenially generate the observed amount of baryogenesis. Such possibilities to realize baryogeneis in the early Universe remain open in the present framework and a detailed analysis {\it e.g.} building a suitable extension of the minimal $L_\mu-L_\tau$ model is kept for a future work.}

\section{Acknowledgements}
AP acknowledges Deep Ghosh, Sk Jeesun, Sougata Ganguly and Tanmoy Kumar for useful discussions and several comments. AP thanks IACS, Kolkata for financial support through Research Associateship. AKS is supported by a postdoctoral fellowship at IOP, Bhubaneswar, India. A.K.S. would like to acknowledge NPDF grant PDF/2020/000797 from the SERB, Government of India, during his stay at IACS where this work was initiated.

\appendix
\section{Comment on preheating}\label{preheat}
In this work we have assumed that the energy density is stored in the inflaton zero momentum mode until it dissipates into radiation at the reheating epoch. However, realistically, during the initial oscillations of the inflaton (much before the perturbative reheating takes place) a fraction of the energy is dissipated both into higher momentum modes of inflaton and to other fields coupled to the inflaton. This process is known as preheating. The dissipation of inflaton energy through preheating results into a relatively longer or shorter radiation like period (from $a_{\phi,\rm rad}$ to $a_{\phi,\rm matter}$) before the matter like oscillation of the inflaton dominates, thereby slightly increasing or decreasing the ratio $\frac{a_{\phi,\rm matter}}{a_{\phi,\rm rad}}$ which we denote by $\exp(\Delta N)$.  

If at the end of preheating, $f$ fraction of the total energy density is in the radiation sector, $f=\frac{\rho_{r}}{\rho_r+\rho_{\rm inf_0}}$ where $\rho_{\rm inf_0}$ is the energy density of zeroth mode of inflaton at the end inflation. Then it is simple to show that there is a change of e-folding number of radiation-like phase by an amount,
\begin{align}
\Delta N=\begin{cases} 
      -\frac{1}{4}\log\frac{1}{1-f}\,, & f\lesssim 0.5 \\
       -\frac{1}{4}\log(2)+\frac{3}{4}\log(\frac{f}{1-f})\, ,& f>0.5
   \end{cases}
\end{align}
As a representative choice of $f=90\%$, we find a small shift in the inflationary e-fold number $N_e$ ({\it i.e.}\,$\Delta N_e \sim$ 0.5) for a particular set of $(m_\phi,\sin\theta)$ corresponding to a constant $\xi$ value. This however does not alter the predicted region (Fig.\,\ref{fig:nsr}) for BP 1 and BP 2 in the $n_s-r$ plane at a noticeable amount. Thus we conclude that even if during preheating, a large fraction of the inflaton energy density gets transferred to the radiation at very early stage of inflaton oscillations, our results remain more or less same.

\section{Decay Rates of Inflaton}\label{sec:DecayExp}
\noindent $\bullet$ $\phi\to \overline{f}f$: 
\begin{equation}
    \Gamma_\phi=\sin^2\theta \times \, \Gamma_h (m_{\phi}),
\end{equation}
where, $\Gamma_h(m_{\phi})$ is the SM Higgs decay width when its mass is $m_\phi$.

\vspace{1mm}
\noindent$\bullet$ $\phi\to Z^\prime\overline{f}f$: 
\begin{align}
        \Gamma_{\phi\rightarrow Z'\overline{f} f} \,=\, \frac{1}{2E_{\rm p}}\int\prod_{i}\frac{\d^3k_i}{(2\pi)^3(2E_{ k_i})}(2\pi)^4\delta^4(P-\sum_{\rm i}K_{i})\,|\mathcal{M}|^2,
\end{align}
where,
\begin{align}
        |\mathcal{M}|^2 \,=\, \frac{16v'^2 g'^4}{(s_1-m_{Z'}^2)^2}\nonumber\times\left[ \frac{s_1-2m_f^2}{2}+3m_f^2 
       + \frac{1}{m_{Z'}^2}\frac{(s_3-m_f^2-m_{Z'}^2)^2(s_2-m_f^2-m_{Z'}^2)^2}{2}\right],
\end{align}
in the limit $g^\prime\gg \sin\theta$. Here, $p$ and $k_i$ stand for the 3-momentum for $\phi$ and a outgoing particle with $i=\{1-3\}$. Furthermore, we define $s_{\rm i}=(P-K_i)^2$ where $P=\{E_p,p\}$ and $K_i=\{E_{k_i},k_i\}$.

\vspace{1mm}
\noindent $\bullet$ $\phi\to Z^\prime Z^\prime$:
\begin{equation}
    \Gamma_{\phi\rightarrow Z'Z'}=\frac{1}{8\pi}\frac{m_{Z'}^4}{m_\phi v'^2}\left( 1-\frac{4m_{Z'}^2}{m_\phi^2}\right)^{1/2}\left(3+\frac{m_\phi^4}{4m_{Z'}^4}-\frac{m_\phi^2}{m_{Z'}^2}\right).
\end{equation}
We have checked also that the energy density of the inflaton oscillations can indeed become negligible with respect to the radiation energy density it is decaying into, even within the radiation like ($w=1/3$) oscillation phase, See Fig.\ref{fig:Boltz} for an example.

\section{Inflaton decay in Case II}\label{sec:DecayInfC2}
{In this section, we would like to inquire whether the reheating gets completed indeed in the $\omega_\phi=1/3$ phase for case II due to a relatively larger decay strength of the inflaton. We use the following set of Boltzmann equation to obtain the evolution pattern of the inflaton and radiation energy densities.
\begin{eqnarray}\label{eq:BeQS}
    \frac{d\rho_\phi}{dt}+3H\rho_\phi(1+\omega_\phi)&=&-\Gamma_\phi\rho_\phi\\
     \frac{d\rho_R}{dt}+4H\rho_R &=&\Gamma_\phi\rho_\phi,
\end{eqnarray}
where $\rho_R$ is the radiation energy density.}
\begin{figure}
    \centering
    \includegraphics[width=.5\textwidth]{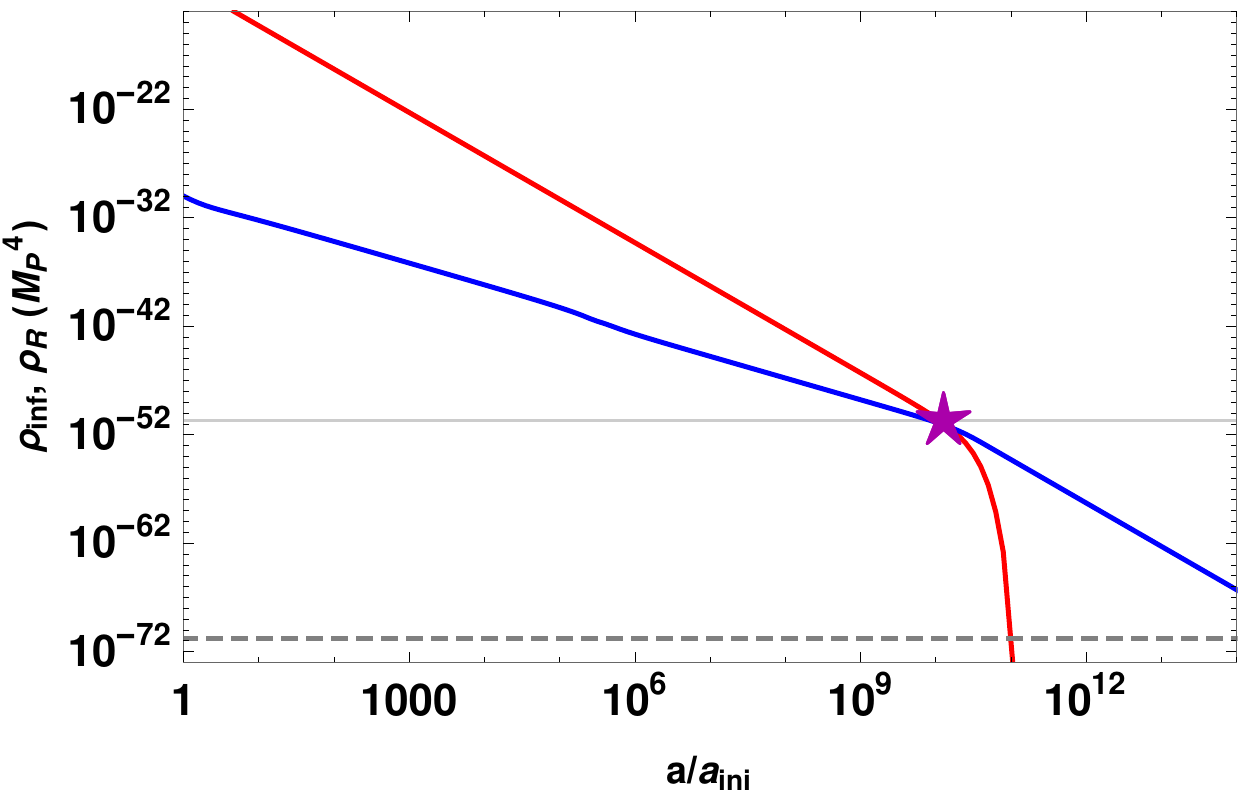}
    \caption{Evolutions of $\rho_\phi$ (red) and $\rho_R$ (blue) after solving Eq.(\ref{eq:BeQS}) for BP 1, considering {\color{magenta}$m_\phi= 0.45$\,GeV}. The gray horizontal dashed line denotes $\rho_\phi=\rho_{\phi,\rm matter}$. The `$\star$' indicates the epoch where the reheating completes and the radiation takes over in the Universe.}
    \label{fig:Boltz}
\end{figure}

 {In Fig.\,\ref{fig:Boltz}, we show the evolutions of $\rho_\phi$ and $\rho_R$ as a function of scale factor considering BP 1. The reheating gets over when $\rho_\phi=\rho_R$ happens as indicated by a star in the Fig.\,\ref{fig:Boltz}. The dashed horizontal line stands for $\rho_\phi=\rho_{\phi,\rm matter}$ which happens at some threshold scale factor. Above this threshold scale factor, the inflaton behaves like a matter. In the present case, we find the cross-over between the $\rho_\phi$ and $\rho_R$ takes place at a scale factor lower than its threshold value indicating that the reheating of the Universe gets completed at $\omega_\phi=1/3$ phase.}

\bibliographystyle{apsrev4-1}
\bibliography{ref}
\end{document}